\documentclass[12pt]{article}
\usepackage{epsfig}
\newskip\humongous \humongous=0pt plus 1000pt minus 1000pt
\def\caja{\mathsurround=0pt}
\def\eqalign#1{\,\vcenter{\openup1\jot \caja
\ialign{\strut $\displaystyle{##}$&$
\displaystyle{{}##}$\hfil\crcr#1\crcr}}\,}
\newif\ifdtup

\def\tr{\mathop{\rm tr}}

\def\pr#1{#1^\prime}

\def\beq{\begin{equation}}
\def\eeq{\end{equation}}

\def\beqn{\begin{eqnarray}}
\def\eeqn{\end{eqnarray}}

\relax

\catcode`\@=11

\@addtoreset{equation}{section}
\def\theequation{\thesection.\arabic{equation}}

\def\@normalsize{\@setsize\normalsize{15pt}\xiipt\@xiipt
\abovedisplayskip 14pt plus3pt minus3pt%
\belowdisplayskip \abovedisplayskip
\abovedisplayshortskip \z@ plus3pt%
\belowdisplayshortskip 7pt plus3.5pt minus0pt}

\def\small{\@setsize\small{13.6pt}\xipt\@xipt
\abovedisplayskip 13pt plus3pt minus3pt%
\belowdisplayskip \abovedisplayskip
\abovedisplayshortskip \z@ plus3pt%
\belowdisplayshortskip 7pt plus3.5pt minus0pt
\def\@listi{\parsep 4.5pt plus 2pt minus 1pt
\itemsep \parsep
\topsep 9pt plus 3pt minus 3pt}}

\@twosidetrue

\relax

\catcode`@=12

\catcode`\@=11

\def\section{\@startsection{section}{1}{\z@}{3.5ex plus 1ex minus
.2ex}{2.3ex plus .2ex}{\large\bf}}
\def\subsection{\@startsection{subsection}{2}{\z@}{3.25ex plus 1ex minus
.2ex}{1.5ex plus .2ex}{\bf}}

\def\thesection{\arabic{section}}
\def\thesubsection{\arabic{section}.\arabic{subsection}}

\def\appendix{\setcounter{section}{0}
\def\thesection{Appendix \Alph{section}:}
\def\thesubsection{\Alph{section}.\arabic{subsection}}
\def\theequation{\Alph{section}.\arabic{equation}}
\def\section{\@startsection{section}{1}{\z@}{3.5ex plus 1ex minus
.2ex}{2.3ex plus .2ex}{\large\bf}}
\def\subsection{\@startsection{subsection}{2}{\z@}{3.25ex plus 1ex minus
.2ex}{1.5ex plus .2ex}{\bf}}}
\def\marksection{\def\@currentlabel{\Alph{section}}}

\def\ps@headings{\def\@oddfoot{}\def\@evenfoot{}
\def\@oddhead{\hbox{}\hfill
\makebox[.5\textwidth]{\raggedright\ignorespaces --\thepage{}--
\hfill {}}}  
\def\@evenhead{\@oddhead}
\def\subsectionmark##1{\markboth{##1}{}}
}

\ps@headings

\catcode`\@=12

\def\r#1{\ignorespaces $^{#1}$}
\def\figcap{\section*{Figure Captions\markboth
{FIGURECAPTIONS}{FIGURECAPTIONS}}\list
{Fig. \arabic{enumi}:\hfill}{\settowidth\labelwidth{Fig. 999:}
\leftmargin\labelwidth
\advance\leftmargin\labelsep\usecounter{enumi}}}
 \relax
\def\tablecap{\section*{Table Captions\markboth
{TABLECAPTIONS}{TABLECAPTIONS}}\list
{Table \arabic{enumi}:\hfill}{\settowidth\labelwidth{Table 999:}
\leftmargin\labelwidth
\advance\leftmargin\labelsep\usecounter{enumi}}}
 \relax
\def\reflist{\section*{References\markboth
{REFLIST}{REFLIST}}\list
{[\arabic{enumi}]\hfill}{\settowidth\labelwidth{[999]}
\leftmargin\labelwidth
\advance\leftmargin\labelsep\usecounter{enumi}}}
 \relax

\catcode`\@=11

\relax

\catcode `@ 11
\def\biblabel#1{\if@filesw\immediate
\write\@auxout{\string\bibcite{#1}{\the\value{\@listctr }}}\fi}
\catcode `@ 12

\newcommand{\ccaption}[2]{
\begin{center}
\parbox{0.85\textwidth}{
\caption[#1]{\small\it {#2}}}
\end{center}    }
\def    \be             {\begin{equation}}
\def    \ee             {\end{equation}}
\def    \ba             {\begin{eqnarray}}
\def    \ea             {\end{eqnarray}}

\def    \=              {\;=\;}
\def    \frac           #1#2{{#1 \over #2}}

\def \ep{\epsilon}
\def \as   {\ifmmode \alpha_s \else $\alpha_s$ \fi}

\def\b0{b_0}

\def \mt   {\ifmmode m_{\rm t} \else $m_{\rm t}$ \fi}

\def \to   {\mbox{$\rightarrow$}}
\newcommand     \MSB            {\ifmmode {\overline{\rm MS}} \else
$\overline{\rm MS}$  \fi}

\newcommand\hepph[1]{{\tt hep-ph/#1}}

\def\dl#1{$$\displaylines{\quad#1}$$}
\def\nl{\hfill\cr\hfill}

\begin{document}
\begin{titlepage}
\nopagebreak
{\flushright{
\begin{minipage}{4cm}
CERN-TH/97-219  \hfill \\
IFUM 589/FT \hfill \\
hep-ph/9709360\hfill \\
\end{minipage}        }

}
\vfill
\begin{center}
{\LARGE  \bf \sc \baselineskip 0.9cm
Next-to-Leading-Order Corrections to\newline
the Production of Heavy-Flavour\newline
Jets  in $e^+e^-$ Collisions

}
\vskip .5cm
{\bf Paolo NASON\footnote{On leave of absence from INFN, Milan, Italy},}
\\
\vskip 0.1cm
{CERN, TH Division, Geneva, Switzerland} \\
\vskip .5cm
{\bf Carlo OLEARI}
\\
\vskip .1cm
{Dipartimento di Fisica, Universit\`a di Milano and INFN, Milan, Italy}
\end{center}
\nopagebreak
\vfill
\begin{abstract}
In this paper we describe the calculation of the process
\mbox{$e^+e^-\,\to\,Z/\gamma\,\to\,Q\overline{Q}+X$},
where $Q$ is a heavy quark, at order $\as^2$.
\end{abstract}
\vskip 1cm
CERN-TH/97-219 \hfill \\
\today \hfill
\vfill
\end{titlepage}

\def\figura#1#2#3
{\begin{figure}[hb]
\begin{center}
\leavevmode\protect\epsfxsize=#1\protect\epsffile{#2.eps}
\protect\caption{#3}
\label{#2}
\end{center}
\end{figure}
}
\newlength{\Largfig}
\Largfig=7.5cm
\def\thp{\phi}
\newcommand\kl{y}
\newcommand\lk{y}
\newcommand\Asq{M}

\section{Introduction}
Radiative corrections to jet production in $e^+e^-$ annihilation
were computed a long time ago \cite{ERT,VGO,FKSS}. These calculations
were, however, performed for massless quarks. In most practical applications
this is sufficient, since, at relatively low energy, the $b$ fraction
is strongly suppressed, and at high energy (i.e. on the $Z$ peak
and beyond) mass effects are  suppressed. Nevertheless
there are several reasons why a next-to-leading-order calculation is desirable.
First of all, at sufficiently high energies, top pairs will be produced
and mass effects there are very likely to be important. A second
reason is to understand the relevance of mass corrections, due to
bottom production, to the determination of $\as$ from event shape variables.
As a third point, quantities such as the heavy-flavour momentum correlation
\cite{NO,NO1}, although well defined
in the massless limit, cannot be computed using the massless results
of refs.~\cite{ERT,VGO,FKSS}.

In this paper we describe a recently completed
next-to-leading-order calculation of the
heavy-flavour production cross section in $e^+e^-$ collisions,
including quark mass effects.
Very recently, two calculations have appeared in the literature
that address the same problem \cite{Rodrigo,Bernreuther,Brandenburg}.
They both use a slicing method in order to deal with infrared divergences.
In our work, we preferred to use a subtraction method,
since, in this way, we do not need to worry about taking the
limit for small cutoff parameters\footnote{Subtraction methods for
the calculation of radiative corrections to $e^+e^-\,\to\,\mbox{jets}$
have been used in refs.~\cite{ERT,YellowBook}, and
they have also been successfully employed in the calculation
of hadronic production processes.}.
We were able to perform a partial comparison of our result with that of
ref.~\cite{Rodrigo}, and found satisfactory agreement.
In the older work of ref.~\cite{Ballestrero},
a calculation of the process $e^+e^-\,\to\, Q\overline{Q}gg$ has been given,
but virtual corrections to the process $e^+e^-\,\to\, Q\overline{Q}g$ were
not included.
In ref.~\cite{Magnea}, the NLO corrections to the production
of a heavy quark pair plus a photon are given, including both real and virtual
contributions. 

The paper is organized as follows. In Section~\ref{sec:generalities}
we give a brief outline of the calculation. In Section~\ref{sec:kinematics}
we introduce our kinematical definitions and conventions. In 
Section~\ref{sec:outlinecalc} we present a somewhat detailed description
of the calculation. In Section~\ref{sec:checks} we describe a few checks
on our result. 
Finally, Section~\ref{sec:conclusion} contains some concluding
remarks.

\section{Generalities}\label{sec:generalities}
\begin{figure}[htb]
\centerline{\epsfig{figure=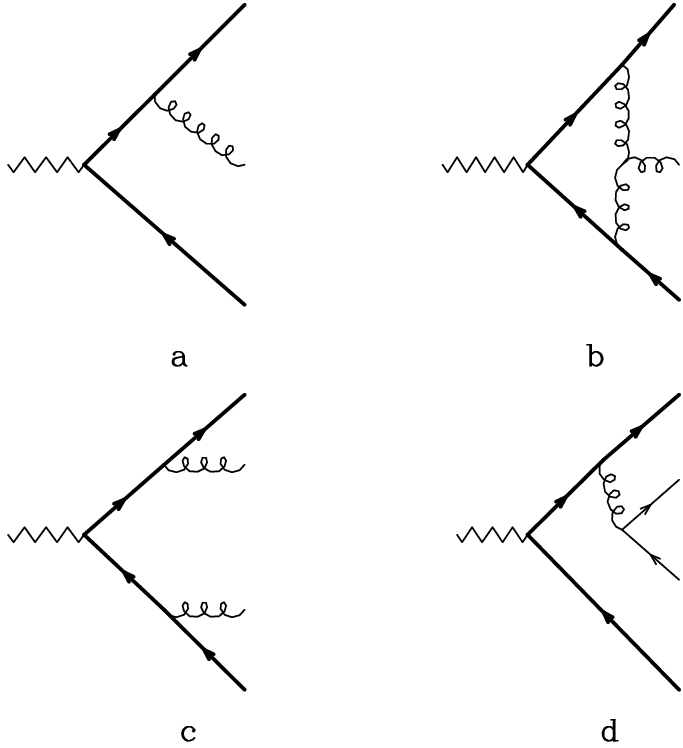,width=0.7\textwidth,clip=}}
\ccaption{}{ \label{fig:graphs}
Some of the diagrams contributing to the process 
$Z/\gamma\,\to\, Q\overline{Q}+X$:
a Born graph (a), a virtual graph (b), a real
emission graph (c) and a real emission graph with light quarks
in the final state (d). }
\end{figure}

We begin by showing in Fig.~\ref{fig:graphs} the Feynman diagrams
for a Born term (a),
a virtual correction term (b) and two real next-to-leading
contributions (c,d).
Next-to-leading corrections arise from the interference
of the virtual graphs with the Born graphs, and from the square
of the real graphs.
Observe that we always deal with the cross section for the production
of the heavy quark pair plus the emission
of at least one extra particle (i.e. a gluon or a quark).
The inclusion of virtual graphs with only a $Q\overline{Q}$ pair in the
final state is not needed if one  computes three-jet-related quantities.
Furthermore, since we deal with unoriented shape variables, the
kinematics of these virtual graphs is fully specified, and in order to
account for them it is enough to include, in the final result, a
two-body contribution normalized in such a way that one obtains the
correct total heavy-flavour cross section at order $\as^2$ (see
\cite{KUHN} and references therein).

Virtual graphs, besides the usual ultraviolet divergences (which are removed
by renormalization), also have infrared and collinear divergences.
These cancel when suitable infrared-safe final-state
variables are considered. Our treatment of the infrared cancellation
is such that the final result is expressed as a partonic event
generator, in which pairs of weighted correlated events are produced.
Shape variables are computed independently for each generated event,
and histogrammed with the corresponding weight.
Infrared-safe shape variables give rise to finite distributions.
No arbitrary cutoff is needed in this calculation in order to implement
the cancellation of virtual and real infrared divergences,
since this cancellation takes place between the two correlated events.
Therefore, one does not have to worry about taking the limit for a vanishing
soft cutoff. This method is similar to the one of ref.~\cite{YellowBook},
which was used there to compute a large class of shape variable distributions
for the LEP experiments.

At next-to-leading order, several complications arise that must be considered.
In fact, heavy flavours may also be produced by a gluon splitting
mechanism, and diagrams with four heavy quarks in the final state are also
present. Interferences between gluon splitting and direct production
should also be considered.
It is useful, however, to separate the various contributions
in the following way. We examine each contribution in terms
of cut Feynman graphs, which represent, individually, a single
contribution to the cross section.
We classify the cut graphs according to the following types:
\begin{itemize}
\item[A)]
 Contributions where the electroweak currents in the cut graphs
 are coupled to the same heavy-flavour loop,
 and there is a single $Q\overline{Q}$ pair in the final state.
 These contributions are the most complex from the point of view of
 renormalization and soft and collinear divergences. They constitute the
 hard part of the calculation. They include graphs in which a pair
 of gluons or a pair of light quarks is present in the final state.
 We will call them A-type.
 We show some of them in Fig.~\ref{fig:A-type}. 
\begin{figure}[htb]
\centerline{\epsfig{figure=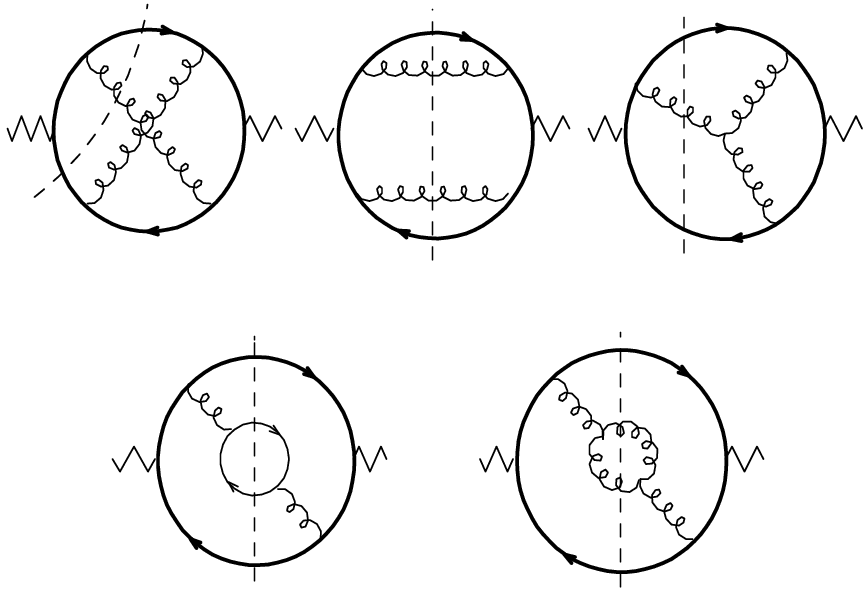,width=0.7\textwidth,clip=}}
\ccaption{}{ \label{fig:A-type}
Some of the diagrams of A-type. Depending upon the cut,
each graph represents a contribution coming from
the square of the four-particle final state
or from the interference between the tree-level graph 
with a virtual correction.}
\end{figure}

\item[B)]
 Contributions where there are two $Q\overline{Q}$ pairs in the final state.
 These include cut graphs with a single heavy-flavour loop coupled to the
 weak currents, graphs with two heavy-flavour loops, one of which is coupled
 to the weak currents, and graphs with two heavy-flavour loops, where
 each loop is coupled to one weak current.
 These contributions are finite, and their computation is a straightforward
 algebraic problem. We will call them B-type.
 We have collected some of them in Fig.~\ref{fig:B-type}. 
\begin{figure}[htb]
\centerline{\epsfig{figure=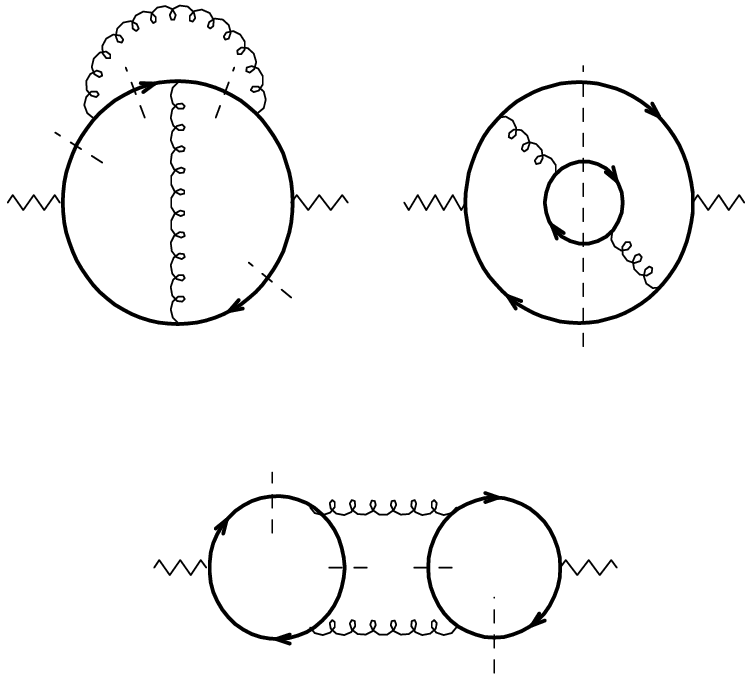,width=0.7\textwidth,clip=}}
\ccaption{}{ \label{fig:B-type}
Some of the diagrams of B-type.}
\end{figure}

\item[C)]
 Contributions where the electroweak currents are coupled to light
 quarks. Also these contributions are finite, and easy to compute.
 The heavy-flavour pair in the final state is generated by gluon splitting.
 We will call them C-type.
\item[D)]
 Interference between terms in which the weak current is coupled to
 the heavy quarks and to quarks of different flavours. These terms
  have the structure of Fig.~\ref{fig:lightheavy}.
\begin{figure}[htb]
\centerline{\epsfig{figure=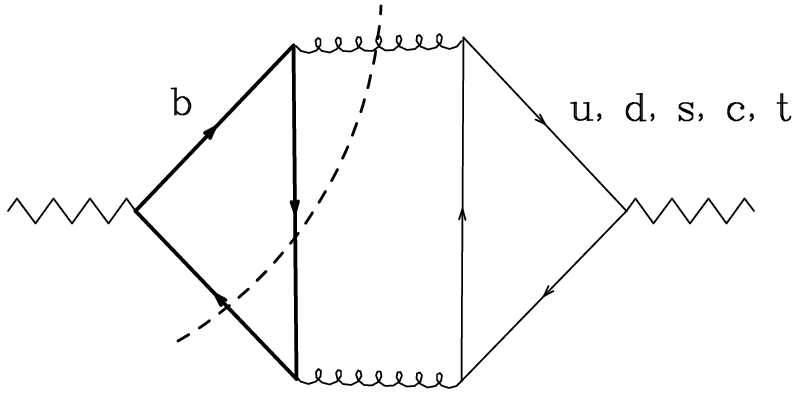,width=0.5\textwidth,clip=}}
\ccaption{}{ \label{fig:lightheavy}
 Structure of cut graphs of D-type.
}
\end{figure}
 By Furry's theorem,
 they must vanish for vector currents. For axial currents, they cancel
 in pairs of up-type and down-type quarks, because they have opposite
 axial coupling. Thus, the up-quark contribution cancels with the down
 quark, and, if the charm mass is neglected, the charm contribution
 cancels with the strange. Only the graph with
 a top quark loop remains. We call these graphs D-type.
\item[E)]
 Graphs with two heavy-flavour loop coupled to the weak current,
 one of which is virtual. We call these graphs E-type. They pair
 naturally with the D-type graphs with the top loop, since in cases
 of practical interest the top loop is also virtual.
\end{itemize}
Most of the following discussion will deal with A-type graphs,
since the other cases are either straightforward, or they have already
been considered in the literature.
For example, graphs of B and C type have been computed in
ref.~\cite{Ballestrero}, and graphs of type D and E have been considered
in ref.~\cite{Hagiwara}.
There is, however, one extra
contribution that should be considered together with the
A-type graphs, that is to say, virtual graphs in which a heavy-flavour
loop corrects the gluon propagator. These graphs are ultraviolet
divergent, and so their inclusion is mandatory if one wants to
have the complete cancellation of ultraviolet divergences after
renormalization.
We will discuss this contribution in detail when we  deal with
renormalization.

A-type graphs contain ultraviolet, soft and collinear divergences
that must be regulated.
Soft divergences arise when, in addition
to the basic $Q\overline{Q}g$ final state, an extra soft gluon is emitted,
giving rise to a real soft divergent contribution.
Collinear singularities arise when the final-state gluon
in the $Q\overline{Q}g$ process undergoes a real (virtual) splitting into
either a $gg$ pair or a $q\bar{q}$ pair. In the first case, the collinear
gluon can also be soft, so that collinear singularities can overlap
with soft singularities.

In order to regularize all singularities we used dimensional regularization.
At first sight, this procedure would seem in conflict with
the presence of the axial coupling. 
In fact, for the class of graphs of A-type,
there is a simple trick to avoid this problem.
First of all, we notice that for unoriented shape variables the axial-vector
interference cannot contribute. In fact, for the three-parton final state
there are not enough momentum vectors to construct an invariant
with an $\epsilon$ symbol. For the four-parton final state one
could in principle
build such an invariant, but the cross section must be symmetric in the
light parton momenta, so that such an invariant cannot survive.
We then consider
the case of a generic vector current coupled to two fermions with
different masses $m_1$ and $m_2$. One can then easily convince oneself
that the case of the axial coupling can be obtained by setting
$m_1=m$ and $m_2=-m$, since one can turn $-m$ into $m$ by a chiral rotation.
This procedure is bound to work if there are no anomalies involved in the
calculation, and this is certainly the case for our A-type graphs.
We will therefore proceed to compute the ${\cal O}(\as^2)$ three-
and four-body contributions in $d=4-2 \epsilon $ dimensions.
We will get a result of the form
\beq
d\sigma = \left(\frac{\as}{2\pi}\right) d\sigma^{(1)}
       + \left(\frac{\as}{2\pi}\right)^2 d\sigma^{(2)}
\eeq
\beq
d\sigma^{(2)} =  \frac{ d\sigma_3^{(2)}}{d\Phi_3} d\Phi_3
+ \frac{ d\sigma_4^{(2)}}{d\Phi_4} d\Phi_4
\eeq
where the suffix $3$ and $4$ refers to the three- and four-body contributions.
The ultraviolet, collinear and soft singularities will manifest themselves
as single and double poles in $1/\ep$ 
in the three-body contribution, and as singularities
arising from the phase-space integration in the four-body contribution.
We will call $q$ the total incoming invariant momentum
and $p$, $p^\prime$ the momenta of the outgoing heavy quark and antiquark.
Furthermore, we introduce the variables
\beq
x_1=\frac{2q\cdot p}{q^2},\quad\quad x_2=\frac{2q\cdot p^\prime}{q^2},
\quad\quad \kl=\frac{(q-p-p^\prime)^2}{q^2}\,.
\eeq
Here $y$ characterizes the mass of the light system accompanying the
heavy-quark pair. Thus, for Born and virtual graphs we will always
have $y=0$.

In addition
\beqn
d\Phi_3 &=& dx_1\,dx_2\,J_3(x_1,x_2)
\nonumber \\
d\Phi_4 &=& dx_1\,dx_2\,d\kl\,d^2 Y\,J_4(x_1,x_2,\kl,Y)\,,
\eeqn
where $Y$ represents the other two variables that are necessary to describe
the four-body final state. In order to implement the cancellation
of the soft and collinear singularities, we now imagine to compute
some physical quantity $G$, function of the final-state
variables. The reader may think of $G$ as  the combination
of theta functions that characterize a histogram bin for some
infrared-safe shape variable.
In general the definition of $G$ will be specified  for any number
of particles in the final state.
Since we are only dealing with three- and four-parton final states,
as far as we are concerned here, $G$ is characterized by only two
functions, $G^{(3)}(x_1,x_2)$ and $G^{(4)}(x_1,x_2,\kl,Y)$.
Soft and collinear finiteness of $G$ will require that
\beq\label{eq:irsafeness}
\lim_{\kl\to 0} G^{(4)}(x_1,x_2,\kl,Y)=G^{(3)}(x_1,x_2)\,.
\eeq
We will have
\beqn
\int  d\sigma^{(2)}  G &=& \int dx_1\,dx_2\,J_3(x_1,x_2)
\frac{ d\sigma_3^{(2)}}{d\Phi_3}
  \,G^{(3)}(x_1,x_2)\nonumber \\
&+& 
\int dx_1\,dx_2\,d\kl\,d^2 Y\,
J_4(x_1,x_2,\kl,Y)\, \frac{d\sigma_4^{(2)}}{d\Phi_4}\, G^{(4)}(x_1,x_2,\kl,Y)
\,,
\eeqn
where each term on the right-hand side contains soft and collinear
divergences
that cancel in the sum.
This formula will be rewritten in the following way:
\begin{displaymath}
\int  d\sigma^{(2)} G = 
 \int dx_1 dx_2 \,G^{(3)}(x_1,x_2) \Bigg\{
\frac{ d\sigma_3^{(2)}}{d\Phi_3} J_3(x_1,x_2)   + 
\int d\kl\,d^2 Y \frac{d\bar{\sigma}_4^{(2)}}{d\Phi_4}
J_4(x_1,x_2,\kl,Y)
\Bigg\}
\nonumber
\end{displaymath}
\beq
+ \int dx_1\,dx_2\,d\kl\,d^2 Y\,J_4(x_1,x_2,\kl,Y)
\Bigg\{\frac{ d\sigma_4^{(2)}}{d\Phi_4}
 G^{(4)}(x_1,x_2,\kl,Y) - \frac{ d\bar{\sigma}_4^{(2)}}{d\Phi_4}
 G^{(3)}(x_1,x_2) \Bigg\} 
\label{eq:subtractionmethod}
\eeq
where $\bar{\sigma}^{(2)}_4$ is chosen  in such a way that it
has the same soft and collinear singular part as $\sigma^{(2)}_4$,
or, schematically
\beq \label{eq:samesing}
\lim_{\kl\to 0} \,\frac{\displaystyle \frac{ d\bar{\sigma}_4^{(2)}}{d\Phi_4} }
{\displaystyle\frac{ d \sigma_4^{(2)}}{d\Phi_4} }\,=\,1\,.
\eeq    
The first term of eq.~(\ref{eq:subtractionmethod}) can be computed
analitically.  The $1/\ep$ single and double poles present in
$d\sigma^{(2)}_3/d\Phi_3$ all cancel with the poles arising from the
$dy\,d^2Y$ integration of $d\bar{\sigma}_4^{(2)}/d\Phi_4$, and thus this
term is finite.

The  second term in eq.~(\ref{eq:subtractionmethod}), because of
eqs.~(\ref{eq:irsafeness}) and (\ref{eq:samesing}), has no soft and collinear
singularities, and thus can be evaluated directly in four
dimensions\footnote{
Observe that both eq.~(\ref{eq:irsafeness}) and
eq.~(\ref{eq:samesing}) must be satisfied in $d$ dimensions in order for
this argument to apply.}.
It is easy to see how the computation of this term can be implemented
numerically. Assuming for simplicity that we can generate four-body
configurations uniformly in the four-body phase space, to each four-body
configuration $x_1,x_2,y,Y$ we associate two events: one four-body events
with kinematics $x_1,x_2,y,Y$ and weight $d\sigma^{(2)}_4/d\Phi_4$,
and one three-body event with kinematics $x_1,x_2$ (and $y=0$),
and weight $-d\bar{\sigma}_4^{(2)}/d\Phi_4$. The
computation of a shape variable using the above scheme reproduces
exactly the second term of eq.~(\ref{eq:subtractionmethod}).
\section{Kinematics}\label{sec:kinematics}
\def\({\left(}
\def\){\right)}
\def\ga{\left\{}
\def\gc{\right\}}
\def\aq{\left\[}
\def\cq{\right\]}
\def\nbar{\bar{n}}
\def\e{\epsilon}
\def\a{\alpha}
\def\b{\beta}
\def\bs{\bar\beta}
\def\ts{\bar t}
\def\l{\lambda}
\def\d{\Delta}
\def\g{\gamma}
\def\dpr{\Delta '}
\def\su{\sigma_1}
\def\sd{\sigma_2}
\def\st{\sigma_3}
\def\s{\sigma}
\def\lp{\lambda_{+}}
\def\lm{\lambda_{-}}
\def\lpm{\lambda_{\pm}}
\def\th{\theta}
\def\zp{\xi_{+}}
\def\zm{\xi_{-}}
\def\zpm{\xi_{\pm}}
\def\cp{\xi_{+}}
\def\cm{\xi_{-}}
\def\cpm{\xi_{\pm}}
\def\r{\rho}
\def\rop{\rho_{+}}
\def\rom{\rho_{-}}
\def\ropm{\rho_{\pm}}
\def\di{(-)}
\def\so{(+)}
\def\as{\alpha_s}
\def\tol{\longrightarrow}
\def\li#1{\,{\rm Li}_2\left(#1 \right)}
\def\dl#1{$$\displaylines{\quad#1}$$}
\def\nl{\hfill\cr\hfill}
\def\tr{\mathop{\rm Tr}}
\def\ds#1{#1\kern-1ex\hbox{/}}
\def\mod#1{|{\bf #1}|}
\def\mean#1{\langle #1 \rangle}
\def\kperp{k_\perp}
\def\tperp{k_\perp}
\def\gperp{g_\perp}

\subsection{Three-body kinematics}
We consider the following three-body process
\beqn
    e^+\(p_e'\)+ e^-\(p_e\) \;\to\;
 Z/\gamma\,(q) \;\to\; Q(p) + \overline{Q}(p') + g(k)
\eeqn 
where $Q$ is the massive quark, and the momenta of the particles satisfy
\[     
p^2=p'^2 = m^2  \hspace{1cm}  k^2 = 0\,.
\]
Since we are interested in unoriented shape variables, we can express 
the three-body phase space in terms of two variables, which we choose to be
\beq
\label{eq:x1x2}
x_1=\frac{2q\cdot p}{q^2},\quad\quad x_2=\frac{2q\cdot p^\prime}{q^2}\,.
\eeq
Defining in addition
\beq
\label{eq:def_rho}
 \rho = \frac{4 m^2}{q^2}\,,
\eeq
the three-body phase space in $d=4-2\ep$ dimensions takes the form 
\beqn
({\rm PS})^{(3)}&=&  H \int_{\sqrt{\rho}}^{1}  dx_1 \int_{x_{2-}}^{x_{2+}}
dx_2
\nonumber \\&&
\label{eq:ps3}
 \ga 4\(x_1^2-\rho\) \(x_2^2-\rho\)-\left[
x_g^2-(x_1^2-\rho)-(x_2^2-\rho) \right]^2\gc ^{-\e}
\eeqn
where:
\beqn
\label{eq:H}
 H\!\! &=&\!\! \frac{1}{\Gamma(2-2\e)}\,
\frac{q^2}{2\, (4\pi)^3}\, \( \frac{16 \pi}{q^2} \)^{2\e}
\\ \label{eq:x2pm}
 x_{2\pm}\!\!\!\! &=&\!\! \frac{1}{4(1-x_1)+\rho} \left[ 2\(1-x_1\) \(2-x_1\) +
      \rho\(2-x_1\)  \pm
    2\(1-x_1\)\sqrt{x_1^2-\rho}   \right]\,.\phantom{a}
\eeqn

\subsection{Four-body kinematics}
The four-body processes we are considering  are
\beqn
  e^+\(p_e'\)+ e^-\(p_e\) \;\to\; 
    Z/\gamma\,(q) \;\to\; Q(p) + \overline{Q}(p') + g(k) + g(l)&&
\nonumber \\
  e^+\(p_e'\)+ e^-\(p_e\) \;\to\;
     Z/\gamma\,(q) \;\to\; Q(p) + \overline{Q}(p') + q(k) + \bar{q}(l)\,, && 
\nonumber
\eeqn
where $q$ is the massless quark.
The momenta satisfy
\beq  
l^2 = k^2 = 0  \quad\quad    p^2 = p'^2 = m^2  \,.
\eeq
In the centre-of-mass system of the two massless particles, we have
\beqn
l &=& l_0 \, \bigl( 1,\ldots,\sin\th\sin\thp,
\sin\th \cos\thp,\cos\th \bigr)  \nonumber \\
k &=& k_0 \, \bigl( 1,\ldots,-\sin\th\sin\thp,
-\sin\th \cos\thp,-\cos\th \bigr)
\nonumber \\
p &=& p_0 \left( 1,\ldots,0,0,\sqrt{1-\frac{m^2}{p_0^2}}\right)
\nonumber \\
p' &=& p'_0 \left(1,\ldots,0,\sqrt{1-\frac{m^2}{{p'_0}^2}}\sin\a,
             \sqrt{1-\frac{m^2}{{p'_0}^2}}\cos\a\right)
\nonumber
\eeqn
where the dots indicate $d-3$ equal and opposite components in the expression
for $l$ and $k$, and $d-3$ zeros in the expression for $p$ and $p'$.

To describe the unoriented four-body phase space, we need five independent 
variables, which we choose to be 
\beq
\label{eq:x1x2y}
x_1=\frac{2q\cdot p}{q^2},\quad\quad x_2=\frac{2q\cdot p^\prime}{q^2},
\quad\quad \kl=\frac{(k+l)^2}{q^2},\quad\quad \th, \quad\quad\thp\;.
\eeq
We thus have
\beq\label{eq:p0pp0}
l_0 = k_0 = \sqrt{q^2}\, \frac{\sqrt{\kl}}{2}, \quad\quad
p_0 = \sqrt{q^2}\,\frac{1-x_2-\kl}{2\sqrt{\kl}}, \quad\quad
p'_0 =\sqrt{q^2}\,\frac{1-x_1-\kl}{2\sqrt{\kl}}
\eeq
and
\beq
\label{eq:cosalfa}
\cos\a = \frac{ \kl\, (\rho-x_1-x_2) + (1-x_1)(1-x_2)-\kl^2}{
\sqrt{\( 1-x_1-\kl \)^2-4 \rho\, \kl}\;
\sqrt{\( 1-x_2-\kl \)^2-4 \rho\, \kl}}\,.
\eeq
Setting
\[
  v = \frac{1}{2}(1-\cos\th)
\]
and defining
  \beqn
  \label{eq:xg_def}
  x_g &=& 2-x_1-x_2  \\
  \kl_{\pm} &=& \frac{1}{4}
  \left[\pm 2\sqrt{x_1^2-\rho}\sqrt{x_2^2-\rho}+ x_g^2
  -(x_1^2-\rho)-(x_2^2-\rho)  \right]
  \nonumber \\ \label{limitsdef}
  x_{1+} \!\!&=& 2-\frac{2-\rho}{2-\sqrt{\rho}}
  \eeqn
the four-particle phase space in $d=4-2\e$ dimensions is
\beqn
({\rm PS})^{(4)} &=& \frac{q^2} {(4\pi)^2 \,
\Gamma(1-\e)}\( \frac{4\pi}{q^2} \)^{\e} H  \nonumber \\
&\times& \ga
  \underbrace{\int_{\sqrt{\rho}}^1 dx_1  \int_{x_{2-}}^{x_{2+}} dx_2
  \int_0^{\kl_{+}} d\kl}_{\rm{region\ I}}
 + \underbrace{
  \int_{\sqrt{\rho}}^{x_{1+}} dx_1  \int_{\sqrt{\rho}}^{x_{2-}} dx_2
  \int_{\kl_{-}}^{\kl_{+}} d\kl }_{\rm{region\ II}}
\gc
\kl^{-\e}
\nonumber \\
&\times& \ga 4\(x_1^2-\rho\) \(x_2^2-\rho\)-\left[
(x_g^2-4 \kl)-(x_1^2-\rho)-(x_2^2-\rho) \right]^2\gc ^{-\e}
\nonumber \\
\label{eq:ps4}
&\times& \int_0^1 dv \,[v(1-v)]^{-\e} \,\frac{1}{N_{\thp}}
\int_0^\pi d\thp\,(\sin\thp)^{-2\e}
\eeqn
with
\beq
N_{\thp}=\int_0^\pi d\thp\,(\sin\thp)^{-2\e}  = 4^\e \pi
\frac{\Gamma(1-2\e)}{\Gamma^2(1-\e)}
\eeq

\begin{figure}[htb]
\centerline{\epsfig{figure=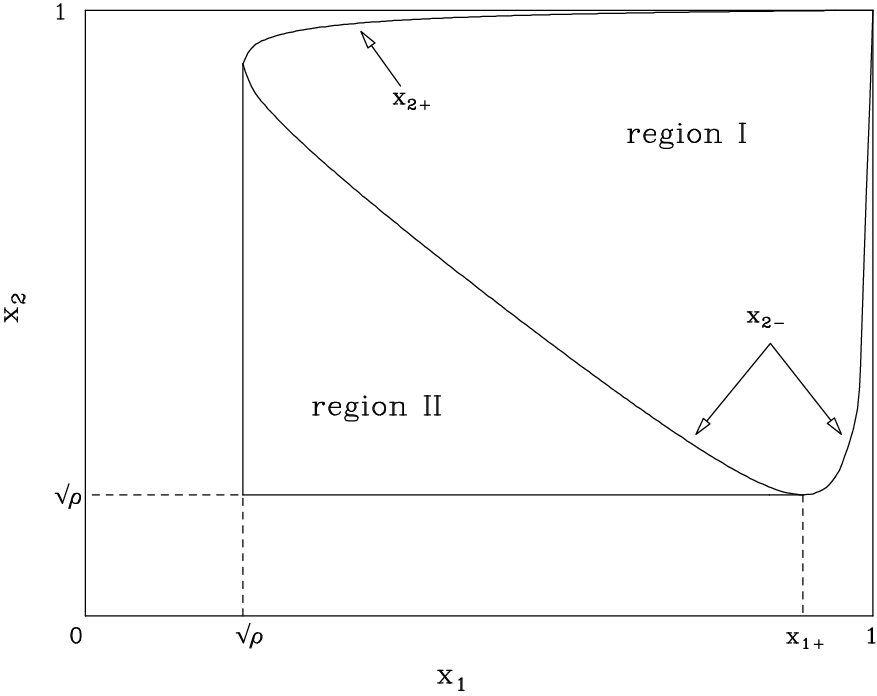,width=0.7\textwidth,clip=}}
\ccaption{}{ \label{phs4}
The two different areas in the $x_1$-$x_2$ plane  correspond to the
{\rm region\ ~I} and to the {\rm region\ ~II} 
of eq.~(\protect{\ref{eq:ps4}}) }
\end{figure}

As can be seen from Fig.~\ref{phs4},
the integration region is split into two parts, one of which (region I) is
characterized by the same $x_1$ and $x_2$ integration limits
as the three-body phase space. In this region, the variable $y$ can reach 
0, and therefore collinear and soft divergences arise.

Sometimes we will need an analogous set of final-state variables,
in which the role of $p$ and $p'$ are interchanged.
The variable $y$ remains the same, $x_1$ and $x_2$ are exchanged, and
the other two variables, denoted by $v'$ and $\thp'$,
are related to $v$ and $\thp$ by the equations
\beq
\label{eq:ppexch}
\eqalign{
v'= \frac{1}{2}\Bigl(1-\cos\alpha
 -(\sin\theta\cos\thp\sin\alpha-2v\cos\alpha)\Bigr)   
\cr
\cos\thp' = \frac{1-\cos\alpha-2\,
(v-v'\cos\alpha)}{2\sin\alpha\,\sqrt{v'(1-v')}}
\,.}
\eeq
Exchanging instead the roles of $l$ and $k$ brings about the following
transformations:
\beq
\label{eq:klexch}
v\,\to\,1-v,\quad \thp\,\to\, \pi+\thp,\quad  v'\,\to\, 1-v',\quad 
\thp'\,\to\, \pi+\thp'\,.
\eeq

\section{Outline of the calculation}
\label{sec:outlinecalc}
The amplitude for the process can be written (up to an irrelevant phase)
\newcommand\sq{{\rm \scriptscriptstyle Q}}
\beqn &&
\bar{u}(p_e) \Bigg[
 g_Z^2 \frac{-g_{\mu\nu}}{q^2-M_{\rm Z}^2+i\Gamma M_{\rm Z}}
\left(v_e\gamma^\mu-a_e\gamma^\mu\gamma^5\right)\langle 0|
J^\nu_V(0) v_\sq-J^\nu_A(0) a_\sq |f \rangle \nonumber \\ &&
\mbox{}\hspace{2cm}+g^2\frac{-g_{\mu\nu}}{q^2}(c_e\gamma^\mu)\langle 0|
J^\nu_V(0) c_\sq |f \rangle \Bigg] v(p_e') \label{epemamp}\,,
\eeqn
where $f$ refers to states with four-momentum $q$.
We use the notation
\beqn
g_{\rm Z}&\equiv&\frac{g}{2\sin \theta_{\rm W} \cos \theta_{\rm W} }
\nonumber \\
v_i&\equiv& T_{3i}-2c_i\sin^2\theta_{\rm W}
\nonumber \\
a_i&\equiv& T_{3i}  \nonumber
\eeqn
where $g$ is the electromagnetic coupling, $T_{3i}$ is the third
component of the (left) isospin of fermion $i$,
$c_i$ is its electric charge in units of the positron charge and
$\theta_{\rm W}$ is the Weinberg angle.
Since we are interested in unoriented events, and following the assumptions
described in  Section~\ref{sec:generalities}, we can neglect
the axial-vector interference in the square of the amplitude.
From eq.~(\ref{epemamp}) we get the following cross section,
averaged over the incoming electron beam direction
\beqn 
d\sigma &=& \frac{N_c\,4\pi\alpha^2}{3 q^2}
\Bigg\{dT_A
\left[\rho_2(q^2)\,(v_e^2+a_e^2)\,a_\sq^2\right]
\nonumber \\ &&
+ \,dT_V
\left[\rho_2(q^2)\,(v_e^2+a_e^2)\,v_\sq^2+c_e^2 \,c_\sq^2-2\,\rho_1(q^2)\,
 v_e\, v_\sq \,c_e \, c_\sq  \right]\Bigg\}  \nonumber
\eeqn
where $\alpha$ is the electromagnetic coupling constant and $N_c=3$ is 
the number of colours.
In addition
\beqn
\rho_1(q^2) &=& \frac{1}{4\sin^2\theta_{\rm W}\cos^2\theta_{\rm W}}\;
\frac{q^2\,(m_Z^2-q^2)}{(m_Z^2-q^2)^2+m_Z^2\,\Gamma_Z^2}\;,\nonumber \\
\rho_2(q^2) &=& \left(\frac{1}{4\sin^2\theta_{\rm W}
               \cos^2\theta_{\rm W}}\right)^2
\frac{q^4}{(m_Z^2-q^2)^2+m_Z^2\,\Gamma_Z^2} \;\nonumber .
\eeqn
We have also defined
\beqn
dT_{V/A}&=&\sum_n {\cal M}^{(f_n)}_{V/A} d\phi_n \nonumber \\
\label{calMdef}
{\cal M}^{(f_n)}_{V/A}&=&\frac{2\pi}{N_c\, q^2}
 \left(-g_{\mu\nu}+\frac{q_\mu q_\nu}{q^2}\right)
\langle 0| J^\mu_{V/A}(0)|f_n\rangle\langle f_n|J^\nu_{V/A}(0)|0\rangle\,,
\eeqn
where $d\phi_n$ represents the $n$-body phase space, and $f_n$ represents
an $n$-body final state.
The $q^\mu q^\nu$ term in the projector in eq.~(\ref{calMdef}) is, of course,
irrelevant for the vector current component, but it should be kept
for the axial current when the quark mass is non-zero.

In the following we will be interested in strong corrections up to the
second order, and into the final states:
 $Q\overline{Q}$, $Q\overline{Q}g$, $Q\overline{Q}gg$
and  $Q\overline{Q}q\bar{q}$. We will use the following simplified notation:
\begin{itemize}
\item
${\cal M}^{(2)}_{V/A}$ for the $Q\overline{Q}$ Born term  
\item
${\cal M}^{(b)}_{V/A}$ or ${\cal M}_b$
to indicate the three-body $Q\overline{Q}g$, ${\cal O}(\as)$ term
\item
${\cal M}^{(v)}_{V/A}$ or ${\cal M}_v$
to indicate the three-body $Q\overline{Q}g$, ${\cal O}(\as^2)$ term
\item
${\cal M}^{(gg)}_{V/A}$ or ${\cal M}_{gg}$
for the four-body $Q\overline{Q}gg$, ${\cal O}(\as^2)$ term
\item
${\cal M}^{(q\overline{q})}_{V/A}$ or ${\cal M}_{q\overline{q}}$
for the four-body $Q\overline{Q}q\overline{q}$, ${\cal O}(\as^2)$ term,
\end{itemize}
and equivalent ones for the $dT_{V/A}$ terms.

We will drop the ${V/A}$ suffix when not referring
specifically to the axial or vector contribution.

\subsection{$Q\overline{Q}$ cross section}
From the amplitude
\beq
\overline {u}(p) \Gamma^\mu_{V/A}  v (p')\,,
\eeq
where $\Gamma^\mu_V=\gamma^\mu$ and $\Gamma^\mu_A=\gamma^\mu\gamma^5$,
we obtain the two-body cross section at zeroth order in $\as$. We get
\beq
{\cal M}_V^{(2)} 
= \frac{2\pi}{N_c\, q^2} N_c\, 4q^2\,\left(1+\frac{\rho}{2}\right)
\,,\quad
{\cal M}_A^{(2)} = 
\frac{2\pi}{N_c\, q^2} N_c\, 4q^2 \beta^2 \, \label{ttwobody}\,,
\eeq
where $\rho$ is defined in~(\ref{eq:def_rho}) and
\beq
 \beta=\sqrt{1-\rho}\,.
\eeq
Multiplying eq.~(\ref{ttwobody}) by the 2-body phase space $\beta/(8\pi)$
we get the zeroth-order total cross section
\beq
T_V^{(2)}=\beta\,\left(1+\frac{\rho}{2}\right)\,,\quad
T_A^{(2)}=\beta^3\,.
\eeq
Thus, our choice for the normalization factor in eq.~(\ref{calMdef})
is such that, in the massless limit, $T_V^{(2)}=T_A^{(2)}=1$.

\subsection{$Q\overline{Q}g$ cross section at order $\as$}
This is obtained starting from the amplitude
\beq
{\cal A}^{\mu\sigma}_{V/A}=\overline {u}(p) \left[ \g^\s
  \frac{\ds{p} +\ds{k}+m}{(p+k)^2-m^2} \Gamma^\mu_{V/A} + \Gamma^\mu_{V/A}
  \frac{\ds{p} -\ds{q}+m}{(p-q)^2-m^2}\g^\s \right]  v (p')
\,.
\eeq
We define
\beq
M^{\sigma\sigma'}_{V/A}
= \left(-g_{\mu\nu}+\frac{q_\mu q_\nu}{q^2}\right) \sum 
{\cal A}^{\mu\sigma}_{V/A}
 {\cal A}^{*\nu\sigma'}_{V/A}
\eeq
where the sum refers to the spin of the fermions in the final state.\newline
The sum over the gluon polarization is
\beq
M_{V/A}=-g_{\sigma\sigma'}\,M_{V/A}^{\sigma\sigma'}\,.
\eeq
We will need $M_{V/A}$  in $d=4-2\e$ dimensions. We have
\beqn
M_V &=&
8 \frac{x_1^2+x_2^2}{(1-x_1)(1-x_2)} +\frac{16}{(1-x_1)^2(1-x_2)^2}
\(\frac{m^2}{q^2}\) \Bigl[ 2\,x_1x_2(x_1+x_2) \nonumber\\&&
-3\,(x_1^2+x_2^2)
-8\,(1-x_1)(1-x_2)+2  \Bigr] - \frac{32}{(1-x_1)^2(1-x_2)^2}
\(\frac{m^2}{q^2}\)^2 x_g^2 \nonumber \\&&
-\frac{16\e}{(1-x_1)(1-x_2)}\Biggl[
x_1^2+x_2^2+(1-x_1)(1-x_2)+x_g-1 \nonumber \\&&
 - \(\frac{m^2}{q^2}\) \frac{x_g^2}
{(1-x_1)(1-x_2)} 
\Biggr]+ \frac{8\e^2}{(1-x_1)(1-x_2)} x_g^2\,,
\eeqn
and
\beqn
M_A &=&
8 \frac{x_1^2+x_2^2}{(1-x_1)(1-x_2)} +\frac{16}{(1-x_1)^2(1-x_2)^2}
\(\frac{m^2}{q^2}\) \Bigl[-12 (x_1+x_2-2x_1x_2) \nonumber \\&&
-11 x_1 x_2 (x_1+x_2) + 8(x_1^2+x_2^2) 
+ x_1^3(x_2-1)+ x_2^3(x_1-1)\nonumber \\&&
+2 x_1^2 x_2^2 
+ 4\Bigr] (1-\e)
 + \frac{64}{(1-x_1)^2(1-x_2)^2}
\(\frac{m^2}{q^2}\)^2 x_g^2 \,(1-\e)\nonumber \\&&
+  \frac{8\e^2}{(1-x_1)(1-x_2)} x_g^2\,,
\eeqn
where $x_1,x_2$ and $x_g$ are defined by~(\ref{eq:x1x2y}) 
and~(\ref{eq:xg_def}).
The three-body, order-$\as$ cross section is given by
\[
{\cal M}^{(b)}_{V/A}=\frac{2\pi}{q^2} \; C_F\, g_s^2 \mu^{2\e}
\, M_{V/A}\,,
\]
where $\mu$ is the mass  parameter of dimensional regularization
and $C_F = \frac{N_c^2 -1}{2N_c} = \frac{4}{3}$
for an SU(3) gauge theory.

We introduce a unit, purely space-like vector $j$ lying in the event plane
(i.e. the plane defined by $\vec{p}$, $\vec{p^\prime}$ and $\vec{k}$),
and orthogonal to $k$
\beq
  j\cdot q=0\,,   \hspace{1cm} j\cdot k = 0\,, \hspace{1cm}  j^2 = -1\,.
\eeq
$M^{\sigma\sigma'}_{V/A}$ has the general form
\beq \label{Mform}
M^{\sigma\sigma'} _{V/A}= M^\perp_{V/A} g^{\sigma\sigma'} +
M^j_{V/A} j^\sigma j^{\sigma'} + \mbox{terms involving $q$ or $k$}\;.
\eeq
In the following, we will need $M^j_{V/A}$, but only in four dimensions
\beqn
M^j_{V/A}&=&  \frac{2\, c_{V/A}}{(1-x_1)^2(1-x_2)^2} \biggl[ 4x_1x_2(x_1+x_2)
 -\rho (x_1+x_2)^2 -4(x_1^2+x_2^2)  \nonumber \\&&
-12 x_1 x_2+ 4(\rho+2)(x_1+x_2)-4(\rho+1) \biggr]\,,
\eeqn
where 
\beq
   c_V = \rho + 2   \hspace{1.5cm}  c_A = -2(\rho-1)\,.
\eeq
We also define, consistently with our previous notation
\beq
\label{eq:Mcalssp}
{\cal M}^{(b)\s\s'}_{V/A} = \frac{2\pi}{q^2} \; C_F\, g_s^2 \mu^{2\e}
\, M^{\s\s'}_{V/A}\,,\hspace{1.5cm} 
{\cal M}^{(b)\perp/j}_{V/A} = \frac{2\pi}{q^2} \; C_F\, g_s^2 \mu^{2\e}
\, M^{\perp/j}_{V/A}\,.
\eeq
In the cases when the $V/A$ suffix needs not be specified,
we will simply write ${\cal M}_b^{\s\s'}$, ${\cal M}_b^\perp$ and
${\cal M}_b^j$. 

\subsection{Virtual contributions}
Corrections to the three-jet decay rate to order $\as^2$
come from the interference of the one-loop graphs
with the tree-level ones. These terms have been computed in $d=4-2\ep$
dimensions. The algebra has been carried out in a straightforward
way, using a MACSYMA program, which reduces the original Feynman graphs
to a linear combination of scalar, one-loop integrals. The scalar
integrals have been computed analytically. Their values are listed
in Appendix~\ref{sec:virtual}.
Loop corrections to on-shell external lines require particular attention.
First of all, gluon and light fermions self-energy corrections to external
gluon lines vanish in dimensional regularization.
Only the corrections coming from a heavy-flavour loop need be considered.
We proceed as follows. We compute the self-energy correction for
a gluon propagator of small virtuality.
We obtain, for the corrected propagator,
\beq
\frac{-i g_{\mu\nu}}{k^2} - N_\ep T_F g_s^2 \left(\frac{\mu^2}{m^2}\right)^\e
\frac{4}{3\ep}
   \frac{-i (g_{\mu\nu}-k_\mu k_\nu/k^2)}{k^2} + \frac{{\cal O}(k^2)}{k^2}\,,
\eeq
where
\beq
\label{eq:def_Nep}
N_\e = \frac{(4\pi)^\e} {(4\pi)^2}\Gamma(1+\e)
\eeq
and the colour factor $T_F = 1/2$.
 
From this equation we immediately infer that the contribution
to ${\cal M}_v$ coming from the
self-energy correction to the external gluon line amounts
to
\beq
 - N_\ep T_F g_s^2 \left(\frac{\mu^2}{m^2}\right)^\ep
\frac{4}{3\ep} \times {\cal M}_b\;.
\eeq
A similar consideration applies to the self-energy corrections to
heavy-flavour external lines. In this case one finds
\beq \label{ggpropcorr}
\frac{i}{\ds{p}-m}- N_\ep C_F g_s^2 \left(\frac{\mu^2}{m^2}\right)^\e
\left(\frac{3}{\ep}+4\right)\frac{i}{\ds{p}-m}-\frac{i}{\ds{p}-m}i\delta m
\frac{i}{\ds{p}-m} + \frac{{\cal O}(p^2-m^2)}{\ds{p}-m} \,,
\eeq
where
\beq
\delta m = N_\ep C_F g_s^2 
\left(\frac{\mu^2}{m^2}\right)^\e\,m\,\left(\frac{3}{\ep}+4\right)\,.
\eeq
The infinite mass correction should be removed by the mass counterterm.
We define the Feynman rule for the mass counterterm to be given
by an insertion of $-i m_c$ in the fermion propagator,
where
\beq
m_c=-\delta m = - N_\ep C_F g_s^2 
\left(\frac{\mu^2}{m^2}\right)^\e\,m\,\left(\frac{3}{\ep}+4\right)\,.
\eeq
This precisely cancels the $\delta m$ term in eq.~(\ref{ggpropcorr}), so
that the pole of the propagator is not displaced by radiative corrections,
and $m$ corresponds to the pole mass definition.
Thus, the effect of the fermion self-energy correction to an external
line, including the effect of the mass counterterm, is given by
\beq
- N_\ep C_F g_s^2 \left(\frac{\mu^2}{m^2}\right)^\e
\left(\frac{3}{\ep}+4\right)\times {\cal M}_b\,.
\eeq
To complete the computation of virtual corrections, the diagrams
with a mass counterterm insertion in internal fermion lines should also
be included. After that, charge renormalization is all that is needed,
since we are computing a physical cross section.
We carry out the charge renormalization in the mixed scheme of ref.~\cite{CWZ},
in which the light flavours $n_{\rm lf}$ are subtracted in the \MSB\ scheme, 
while the heavy-flavour loops  are subtracted at zero momentum. 
In this scheme
the heavy flavour decouples at low energy. The prescription for charge
renormalization in this scheme is
\beq\label{eq:charge_ren}
\as\; \to \; \as \Biggl\{ 1
+ g_s^2 N_\e
\left[ \( \frac{4}{3\e}T_F\,n_{\rm lf} -\frac{11}{3\e}C_A \)
 + \(\frac{\mu^2}{m^2}\)^\e
\frac{4}{3\e}T_F \right]\Biggr\}\;,
\eeq
where $C_A = N_c = 3$ for an SU(3) gauge theory.
It  amounts to adding the following correction to our virtual term
\beq
 g_s^2 N_\e
\left[ \( \frac{4}{3\e}T_F\,n_{\rm lf} -\frac{11}{3\e}C_A \)
 + \(\frac{\mu^2}{m^2}\)^\e
\frac{4}{3\e}T_F \right]\times  {\cal M}_b\;.
\eeq
Observe that in this scheme the term corresponding to the heavy-flavour
loop compensates exactly the self-energy correction to the external
gluon line, coming from the heavy-flavour loop. This is easily understood:
the final-state gluon is on the mass shell, so it is effectively
renormalized at zero momentum by the heavy quark loop, and thus decoupling
applies.
We can now resume the combined effect of external line corrections
and renormalization to be included with ${\cal M}_v$
\beq
N_\ep  g_s^2  \left(\frac{\mu^2}{m^2}\right)^\ep \Bigg\{
- 2 C_F
\left(\frac{3}{\ep}+4\right)
+ \( \frac{4}{3\e}T_F\,n_{\rm lf} -\frac{11}{3\e}C_A \)
\left(\frac{\mu^2}{m^2}\right)^{-\ep}
 \Bigg\} \times {\cal M}_b\;.
\eeq
The factor of 2 in front of the fermion external line corrections is
to account for the two fermion lines.

\subsection{Soft and collinear limit of the
$Q\overline{Q}gg$ and  $Q\overline{Q}q\bar{q}$ \\ cross sections}
\label{sec:sc_limit}
Here we derive an expression for the singular
part of the four-body cross section, valid in both the collinear and the soft
limit. These limits are both characterized by $\lk\,\to\, 0$, except that
in the soft limit, at the same time $v\,\to\, 0$ ($l\,\to\, 0$)
or $v\,\to\, 1$ ($k\,\to\, 0$). 
We will focus our discussion on the $Q\overline{Q}gg$ final state.
The other process $Q\overline{Q}q\overline{q}$ is much simpler, since only
collinear singularities are present there.
Since the same formulae apply irrespective of the vector or axial case,
we will always drop the $V/A$ suffix.

We begin with the soft singularities of ${\cal M}_{gg}$. They are given by
eq.~(\ref{eq:lsoft}), which we now rewrite 
\beqn
{\cal M}_{gg}^{\rm soft} &\sim& g_s^2\,\mu^{2\e}
\Bigg\{
 C_A
\left[\frac{p\cdot k} {(p\cdot l) \, (k\cdot l)} +
\frac{p'\cdot k} {(p'\cdot l) \, (k\cdot l)} \right]
+2 \(C_F-\frac{C_A}{2}\) \frac{p\cdot p'}{(p\cdot l)\,(p'\cdot l)}
\nonumber \\
\label{eq:lsoff}
&& \hspace{2cm}
{}-C_F \left[ \frac{m^2}{(p\cdot l)^2} + \frac{m^2}{(p'\cdot l)^2} \right]
+(k\leftrightarrow l)
 \Bigg\}\times {\cal M}_b\,.
\eeqn
From Section~\ref{sec:kinematics}, we can derive an approximation of
the scalar products in the limit of $l$ soft
\dl{
 \frac{p\cdot k}{(p\cdot l)(k\cdot l)} \sim \frac{2h}{q^2}\,
 \frac{1}{\kl\,[\kl+h\,v]} \equiv E_{p,k;l}(x_1,x_2,y,v)
\hfill}
\dl{
 \frac{p\cdot p'}{(p\cdot l)(p'\cdot l)} \sim
 \frac{K}{m^2} \, \frac{1}{\kl+h\,v}\, \frac{1}{\kl-c
 \cos\thp\sqrt{\kl}\sqrt{v} + g\,v} \equiv E_{p,p';l}(x_1,x_2,y,v,\thp)
\hfill}
\dl{
 \frac{m^2}{(p\cdot l)^2} \sim \frac{4h}{q^2}\,
 \frac{1}{[\kl+h\,v]^2} \equiv E_{p,p;l}(x_1,x_2,y,v)\;,
\hfill}
where
\dl{
 h = \frac{q^2}{m^2}\(1-x_2\)^2   \hfill \cr \quad
 a= \frac{2}{(1-x_1)(1-x_2)} \ga
 x_1+x_2-1-\frac{m^2}{q^2}\left[2+\frac{1-x_2}{1-x_1}+ \frac{1-x_1}{1-x_2}
 \right] \gc \hfill\cr   \quad
 b=\frac{2}{(1-x_1)(1-x_2)} \ga
  x_1+x_2-1-\frac{m^2}{q^2}\left[2+ \frac{1-x_1}{1-x_2}
  \right] \gc  \hfill \cr  \quad
 K = \frac{1-x_2}{1-x_1}\,\frac{4}{b}\left[ x_1+x_2-1-\frac{2
 m^2}{q^2} \right]  \hfill \cr \quad
 c = \frac{2\sqrt{2a}}{b}   \hfill \cr  \quad
 g = \frac{2}{b} \;.\hfill
 }
We will also need analogous formulae in the variables in which the roles
of $p$ and $p'$ are interchanged.
We have
\dl{
 \frac{p'\cdot k}{(p'\cdot l)(k\cdot l)} \sim E'_{p',k;l}(x_1,x_2,y,v')
                       \equiv E_{p,k;l}(x_2,x_1,y,v')
\hfill}
\dl{
 \frac{p\cdot p'}{(p\cdot l)(p'\cdot l)} \sim
  E'_{p,p';l}(x_1,x_2,y,v',\thp') \equiv E_{p,p';l}(x_2,x_1,y,v',\thp')
\hfill}
\dl{
 \frac{m^2}{(p'\cdot l)^2} \sim  E'_{p',p';l}(x_1,x_2,y,v')
 \equiv E_{p',p';l}(x_2,x_1,y,v')\,.
\hfill}
Soft factors for the case when $k$ is soft are instead
obtained from the above using eqs.~(\ref{eq:klexch}).
For example
\beqn
E_{p,l;k}(x_1,x_2,y,v)&=&E_{p,k;l}(x_1,x_2,y,1-v)\;,
\nonumber\\
E_{p,p';k}(x_1,x_2,y,v)&=&E_{p,p';l}(x_1,x_2,y,1-v,\thp+\pi)\;.
\eeqn
We can now write down our approximate soft cross section.
We have
\beqn
{\cal M}_{gg}^{\rm soft}&=& g_s^2\,\mu^{2\e}
\Bigg\{C_A\left[E_{p,k;l}+E'_{p',k;l}
+ E_{p,l;k}+E'_{p',l;k}\right] \nonumber \\ &&
+(C_F-C_A/2) \left[E_{p,p';l}+E'_{p,p';l}+E_{p,p';k}+E'_{p,p';k}\right]
\nonumber \\ &&
-C_F\left[E_{p,p;l}+E'_{p',p';l}+E_{p,p;k}+E'_{p',p';k}\right]\Bigg\}\times
 {\cal M}_b \,.
\eeqn
The soft cross section written in this way is symmetric under the
interchange of $k$ and $l$, and of $p$ and $p'$.

The collinear part of the cross section receives contributions
from both the $gg$ and the $q\overline{q}$ final state. For the $gg$
contribution, according to eq.~(\ref{eq:ggcollfin}), we
can write the collinear part
\beqn &&
 g_s^2\mu^{2\e}
\frac{4 C_A}{q^2 \kl}
\Bigg\{-\left[-2+\frac{1}{z}+\frac{1}{1-z}+z(1-z)\right]g_{\sigma\sigma'}
\nonumber \\ \label{colllim} && \hspace{2cm}
-\; 2 z(1-z)(1-\e)\left[\frac{k_{\perp\s}k_{\perp\s'}}{\kperp^2}
-\frac{g_{\perp\s\s'}}{2-2\e}\right]\Bigg\}\times {\cal M}_b^{\sigma\sigma'}\;,
\eeqn
where $z$ is the momentum fraction of $l$ versus $l+k$ in the
collinear limit. It can be chosen to be equal to $v$ or to $v'$. 

The perpendicular direction refers instead to a direction orthogonal
to $l+k$ in the centre-of-mass system and in the collinear limit.
Using eq.~(\ref{Mform}), the azimuth-dependent term of~(\ref{colllim})
becomes
\beqn &&
g_s^2\,\mu^{2\e}
\frac{4 C_A}{q^2 \kl}
\Bigg\{
-2 z(1-z)(1-\e)\left[{\cal M}^\perp_b+{\cal M}^j_b 
\frac{(k_\perp\cdot j)^2}{k_\perp^2}
-\frac{{\cal M}^\perp_b(2-2\e)-{\cal M}^j_b}{2-2\e}\right]\Bigg\}\;
\nonumber \\ &&  \hspace{2cm}
=g_s^2\,\mu^{2\e}
\frac{4 C_A}{q^2 \kl}
\Bigg\{
-z(1-z){\cal M}^j_b\left[\frac{(k_\perp\cdot j)^2}{k_\perp^2}2(1-\e)
+1\right]\Bigg\}\;.
\eeqn
It is now easy to show that, in the collinear limit,
$(k_\perp\cdot j)^2/k_\perp^2 \to -\cos^2\phi$.

Part of the collinear singularities are already contained
in the soft-limit expression. In fact,
for $\kl\,\to\, 0$ at $v$ fixed, we have
\beq
 E_{p,k;l} \approx E'_{p',k;l} \approx \frac{2}{q^2\kl v}\,,\quad
 E_{p,l;k} \approx E'_{p',l;k} \approx \frac{2}{q^2\kl(1-v)}\,.
\eeq
Thus, the $1/z$ and $1/(1-z)$ terms in the collinear limit
 formula~(\ref{colllim})
should not be included, since they are already present in the
soft term. We thus arrive at the following expression for the collinear
term to be added to the soft term
\beqn
&& {\cal M}_{gg}^{\rm coll}= g_s^2 \mu^{2\e}
\frac{4 C_A}{q^2 \kl}
\Bigg\{ {\cal M}_b \left[\frac{v(1-v)+v'(1-v')}{2}-2\right]
\nonumber \\ && 
+\frac{{\cal M}_b^j}{2} \left[ v(1-v)\left(2(1-\e)\cos^2\phi-1\right)+v'(1-v')
\left(2(1-\e)\cos^2\phi'-1\right)\right]\Bigg\}\phantom{aaa}
\eeqn
An analogous procedure yields an expression for the collinear part
of ${\cal M}_{q\overline{q}}$ (see eq.~(\ref{eq:qqcollfin}))
\beqn
&& {\cal M}_{q\bar{q}}^{\rm coll}= g_s^2 \mu^{2\e}
\frac{4 n_{\rm lf} T_F}{q^2 \kl}
\Bigg\{ {\cal M}_b \frac{1}{4(1-\e)}\left[v'^2+(1-v')^2+v^2+(1-v)^2-2\e\right]
\nonumber \\ &&
\hspace{0.5cm}
-\frac{{\cal M}_b^j}{2} \left[ v(1-v)\left(2\cos^2\phi
-\frac{1}{1-\e}\right)+v'(1-v')
\left(2\cos^2\phi'-\frac{1}{1-\e}\right)\right]\Bigg\}\;.\phantom{aaaa}
\eeqn

The expressions ${\cal M}_{gg}^{\rm coll}$, ${\cal M}_{q\bar{q}}^{\rm coll}$
and ${\cal M}_{gg}^{\rm soft}$ depend
upon $x_1$ and $x_2$ via ${\cal M}_b$ and 
${\cal M}_b^j$.
These expressions are meaningful only if $x_1$ and $x_2$ belong to the
domain of the three-body phase space.
We thus define
\beqn
\widetilde{\cal M}_{gg}&=&
\left({\cal M}_{gg}^{\rm soft} + {\cal M}_{gg}^{\rm coll}\right)\;
\theta_3(x_1,x_2)\,,
\nonumber \\
\label{eq:mtilda}
\widetilde{\cal M}_{q\bar{q}}&=&
 {\cal M}_{q\bar{q}}^{\rm coll}\;
\theta_3(x_1,x_2)\,,
\eeqn
where the $\theta_3$ function is precisely defined to be zero
when $x_1$ and $x_2$ are outside the three-body phase-space region.
More specifically, using the definitions of eqs.~(\ref{limitsdef})
\beq
\theta_3(x_1,x_2)=\theta(1-x_1)\,\theta(x_1-\sqrt{\rho})\,
\theta(x_{2+}-x_2)\,\theta(x_2-x_{2-})\;.
\eeq
We are now in a position to specify the subtraction procedure
outlined in Section~\ref{sec:generalities}.
Our expression for the second-order contribution to an infrared- and
collinear-safe quantity $G$ is given by
\beqn &&
\frac{1}{2}\int {\cal M}_{gg}(x_1,x_2,\kl,v,\phi)\,
 G(x_1,x_2,\kl,v,\phi)\; d\phi_4
\nonumber \\ && \hspace{1cm}+
\int {\cal M}_{q\bar{q}}(x_1,x_2,\kl,v,\phi)\,
 G(x_1,x_2,\kl,v,\phi)\; d\phi_4
\nonumber
+\int {\cal M}_v(x_1,x_2) G(x_1,x_2)\; d\phi_3\;,
\nonumber
\eeqn
where all quantities are computed in $d=4-2\e$ dimensions.
The factor  $1/2$ in front of the $gg$ contribution accounts for the
two identical gluons in the final state.
We rewrite the above expression as
\beqn
&&\frac{1}{2}\int \left({\cal M}_{gg}(x_1,x_2,\kl,v,\phi) G(x_1,x_2,\kl,v,\phi)
     - \widetilde{\cal M}_{gg}(x_1,x_2,\kl,v,\phi) G(x_1,x_2)\right)\; d\Phi_4
\nonumber \\
&&+\int \left({\cal M}_{q\bar{q}}(x_1,x_2,\kl,v,\phi) G(x_1,x_2,\kl,v,\phi)
     - \widetilde{\cal M}_{q\bar{q}}(x_1,x_2,\kl,v,\phi) G(x_1,x_2)\right)\;
 d\Phi_4
\nonumber \\
&&+\int \left({\cal M}_v(x_1,x_2)
+\widetilde{\cal M}_i(x_1,x_2)\right) G(x_1,x_2)
\; d\Phi_3
\eeqn
where we have defined
\beq
\widetilde{\cal M}_i(x_1,x_2)=
\frac{1}{2} \int \widetilde{\cal M}_{gg}(x_1,x_2,\kl,v,\phi)
d\Phi_{4/3}+\int \widetilde{\cal M}_{q\bar{q}}(x_1,x_2,\kl,v,\phi)
d\Phi_{4/3}
\eeq
and $d\Phi_{4/3}$ is defined by
\beq
d\Phi_4\; \theta_3(x_1,x_2) = d\Phi_{4/3}\; d\Phi_3\,.
\eeq
An explicit expression for $d\Phi_{4/3}$ can be obtained
from eqs.~(\ref{eq:ps4}) and (\ref{eq:ps3}). We first notice that
the four-body phase space is almost proportional to the three-body
phase space, except for the ratio
\beq
\left(
\frac{ 4\(x_1^2-\rho\) \(x_2^2-\rho\)-[
(x_g^2-4 \kl)-(x_1^2-\rho)-(x_2^2-\rho)]^2 }
{ 4\(x_1^2-\rho\) \(x_2^2-\rho\)-[
x_g^2-(x_1^2-\rho)-(x_2^2-\rho) ]^2 }
\right)^{-\e}\;=\;1+{\cal O}(\kl\e)\;.
\eeq
On the other hand, terms of order $\kl\e$ can be neglected,
since they cannot generate infrared singularities, because of the $\kl$
factor, and therefore they can only produce terms of order $\e$.
Thus we can write
\beq
d\Phi_{4/3}=  N_\e \,R_\e\, q^2 \,q^{-2\e}
\int_0^{y_+} dy\, y^{-\e} 
 \int_0^1 dv \,[v(1-v)]^{-\e} \,\frac{1}{N_{\thp}}
\int_0^\pi d\thp\,(\sin\thp)^{-2\e}
\eeq
or the analogous one in the $v',\phi'$ variables.
The normalization factor $N_\e$ is defined in~(\ref{eq:def_Nep}), while
\[
R_\e = \frac{1}{\Gamma(1+\e)\Gamma(1-\e)}=
 1-\frac{\pi^2\e^2}{6}+{\cal O}(\e^3)\;.
\]
Since we are free to choose the set of variables we prefer in the 
$d\Phi_{4/3}$ integration,
it is easy to see that the $\widetilde{\cal M}_i(x_1,x_2)$ term reduces to
\beqn 
\widetilde{\cal M}_i(x_1,x_2)&=&g_s^2 \mu^{2\e} \int
 \Bigg\{
\frac{1}{2}\frac{4 C_A}{q^2 \kl} \left(v(1-v)-2\right)
\nonumber \\ &&{}
+\frac{2 n_{\rm lf} T_F}{q^2 \kl} \frac{1}{1-\e}\left[v^2+(1-v)^2 -\e\right]
\nonumber \\
&&{}+
\frac{1}{2}
\bigl[4 C_A E_{p,k;l}
+4(C_F-C_A/2) E_{p,p';l}-4 C_F E_{p,p;l}\bigr]\,\Bigg\}\times {\cal M}_b
\, d\Phi_{4/3}
\nonumber
\eeqn
where the term proportional to ${\cal M}^j$ has been dropped, since it vanishes
in $d=4-2\e$ dimensions, after the azimuthal integration.

Furthermore, the remaining collinear term is easily integrated. We define
\begin{eqnarray}
\label{eq:I_ggcoll}
I_{gg}^{\rm coll} &=& 
\int_0^{y_+} dy\, y^{-\e} 
 \int_0^1 dv \,[v(1-v)]^{-\e} \,\frac{1}{N_{\thp}}
\int_0^\pi d\thp\,(\sin\thp)^{-2\e}
\,\frac{1}{\kl}\, \left[v\,(1-v) -2 \right]
= \nonumber\\
&=& -\frac{1}{\e}\left[1-\e\log
\(\kl_{+}\)\right] \(-\frac{11}{6} - \frac{67}{18}\e\) + {\cal
O}(\e)
\end{eqnarray}
and
\begin{eqnarray}
\label{eq:I_qqcoll}
I_{q\bar{q}}^{\rm coll} &=& 
\int_0^{y_+} dy \, y^{-\e} 
 \int_0^1 dv \,[v(1-v)]^{-\e} \,\frac{1}{N_{\thp}}
\int_0^\pi d\thp\,(\sin\thp)^{-2\e}
\,\frac{1}{\kl}\, \frac{v^2+(1-v)^2-\e}{1-\e}
= \nonumber\\
&=& -\frac{1}{\e}\left[1-\e\log
\(\kl_{+}\)\right] \(\frac{2}{3}+\frac{10}{9}\e\) + {\cal
O}(\e)\;.
\end{eqnarray}
For the integrals of the soft term we define
\[
I_{p,k;l} = q^2 \,
\int_0^{y_+} dy \, y^{-\e} 
 \int_0^1 dv \,[v(1-v)]^{-\e} \,\frac{1}{N_{\thp}}
\int_0^\pi d\thp\,(\sin\thp)^{-2\e}
E_{p,k;l}
\]
and the analogous ones for $I_{p,p;l}$ and $I_{p,p';l}$.
In this way
\[
I_{p,k;l} = 2 h \,I_1  \hspace{1.5cm}
I_{p,p;l} = 4 h \, I_2  \hspace{1.5cm}
I_{p,p';l} = K\, \frac{q^2}{m^2}\, I_3\;,
\]
where the values of $I_1$, $I_2$ and $I_3$ are collected in
Appendix~\ref{appendix:soft_integrals}.

Our final expression for $\widetilde{\cal M}_i(x_1,x_2)$ is therefore
\beqn
\widetilde{\cal M}_i(x_1,x_2)&=& N_\e \,R_\e\,
 g_s^2 \left(\frac{\mu^2}{q^2}\right)^\ep 
 \Bigg\{
2 C_A
\,I_{gg}^{\rm coll}  + 2\, n_{\rm lf}\, T_F\, I_{q\bar{q}}^{\rm coll}
\nonumber \\
&&+
\Bigg[2 C_A I_{p,k;l}
+2(C_F-C_A/2) I_{p,p';l} - 2 C_F I_{p,p;l}\Bigg]\,\Bigg\}\times{\cal M}_b\; .
\nonumber
\eeqn

\section{Checks of the calculation}
\label{sec:checks}
Several checks have been performed to control the correctness of our results.
\begin{enumerate}
\item 
The divergences coming from UV and IR poles all cancel.

\item 
 The full calculation, as $m\,\to\, 0$, agrees with the massless result
  of ref.~\cite{ERT}.

\item
Our four-dimensional matrix elements for the processes $e^+e^-\,\to\,
Z/\gamma\,\to\,Q\overline{Q} g g$ and 
$e^+e^-\,\to\,Z/\gamma\,\to\,Q\overline{Q} Q\overline{Q}$
agree with ref.~\cite{Ballestrero}. Furthermore, the soft and collinear 
limits of the four-body matrix elements for the process
$Z/\gamma\,\to\,Q\overline{Q}$ plus two light partons
are correctly given by formulae~(\ref{eq:mtilda}).
\item
 Near the production threshold, we should recover the Coulomb singularity.
 If $\beta$ is the velocity of the two massive
 quarks in the fermion centre-of-mass system, then (see ref.~\cite{NDE})
 \beq
\label{eq:coulomb}
  d\sigma^{(v)}_{V/A}(x_1,x_2) 
\stackrel{\beta \to 0}{\longrightarrow}
 \frac{\pi^2}{\beta}\(C_F
  -\frac{C_A}{2}\) d\sigma^{(b)}_{V/A}(x_1,x_2)\;.
 \eeq
By evaluating $(p+p')^2$ in the centre of mass of the two massive 
quarks, for small $\beta$, we get
\beq
 (p+p')^2 = \left[ 2\(m + \frac{m}{2} \beta^2 
          + {\cal O}(\beta^4)\) \right]^2 
     = (q-k)^2 = q^2\, (x_1+x_2 -1)\;.
\eeq
Choosing for example $x_1 = x_2$ we have
 \[
  x_1=x_2 =\frac{1}{2}\(1+\rho+\rho\,\beta^2\)\;.
 \]
By letting $\beta$ get smaller and smaller we have checked that the behaviour
of the virtual differential cross section is in agreement 
with eq.~(\ref{eq:coulomb}).
\end{enumerate}

A further check is  described in detail in ref.~\cite{no_fragmentation}.

\section{Conclusion}
\label{sec:conclusion} 
In this paper we have described a next-to-leading-order calculation of the
heavy-flavour production cross section in $e^+e^-$ collisions, including
quark mass effects. Some applications of our calculation have appeared
in the literature \cite{NO1}, \cite{no_fragmentation}.

We have used a subtraction method instead of a slicing method, in
order to avoid having to worry about taking the limit of some small
parameters. We have performed several checks on the correctness of
our results. Among them, the small mass limit of the energy--energy
correlation is of particular significance, since, for this quantity,
some discrepancies among different approaches are still present (see
ref.~\cite{LEP2}).

\appendix
\section{Phase space for four massive quarks}
For completeness, we describe in this appendix the phase space 
for four massive quarks in the final state.
The process is
\beq
  e^+\(p_e'\)+ e^-\(p_e\) \;\to\;
 Z/\gamma\,(q) \;\to\; Q(p) + \overline{Q}(p') + Q(r) + \overline{Q}(r')\;,
\eeq
where
\[     r^2=r'^2=p^2=p'^2=m^2\;.
\]
The four-body phase space
is obtained with a procedure similar to the one given in 
Section~\ref{sec:kinematics}, with the simplification
that now the entire cross section has no infrared or collinear
divergences, so that we can put ourselves directly in $d=4$ dimensions
and  we do not need to divide the phase-space region into two different
pieces.
In the centre-of-mass frame of one heavy quark-antiquark pair we have
\dl{
r = (r_0,\mod{r}\sin\th\sin\thp,
\mod{r}\sin\th \cos\thp,\mod{r}\cos\th)  \hfill
}
\dl{
r' = (r_0,-\mod{r}\sin\th\sin\thp,
-\mod{r}\sin\th \cos\thp,-\mod{r}\cos\th)   \hfill
}
\dl{
p = p_0 \( 1,0,0,\sqrt{1-\frac{m^2}{p_0^2}}\)   \hfill
}
\dl{
p' = p'_0 \(1,0,\sqrt{1-\frac{m^2}{{p'_0}^2}}\sin\a,
             \sqrt{1-\frac{m^2}{{p'_0}^2}}\cos\a\) \;,  \hfill
}
where
\[
y =\frac{(r+r')^2}{q^2}  \Longrightarrow r_0 = \sqrt{q^2}\,\frac{\sqrt{y}}{2}
\]
and $p_0$, $p'_0$ and $\cos\a$ are given by~(\ref{eq:p0pp0}) 
and~(\ref{eq:cosalfa}), while $\mod{r}=\sqrt{r_0^2-m^2}$.

The four-body phase space is given by
\beq
\label{eq:ps4m}
({\rm PS})^{(4)} =  \frac{q^4}{\(4\pi\)^6} 
\int_{\rho}^{\bar{y}_{+}} dy \sqrt{1-\frac{\rho}{y}}
\int_{\sqrt{\rho}}^{\bar{x}_{1+}}  dx_1
\int_{\bar{x}_{2-}}^{\bar{x}_{2+}} dx_2  \int_0^1 dv \int_0^{2\pi} d\thp\;,
\eeq
where
\beqn
  \bar{y}_{+} &=& \(1-\sqrt{\rho}\)^2  \nonumber \\
  \bar{x}_{1+} \!\!&=& 1-y-\sqrt{\rho\,y} \nonumber \\
  \bar{x}_{2\pm} \!\! &=& \frac{1}{4(1-x_1)+\rho} \Bigl[
  (2-x_1)(2+\rho-2y-2x_1)  \nonumber \\
    && \pm \,2\, \sqrt{\(x_1^2-\rho\) \left[(x_1-1+y)^2-\rho\,y \right]}
            \Bigr]\;.
\eeqn
A statistical factor $1/(2!2!)=1/4$ must be supplied to~(\ref{eq:ps4m}),
because of the
presence of two pairs of identical particles in the final state.

\section{Collinear limit for $g\to gg$ splitting}
\begin{figure}[htb]
\centerline{\epsfig{figure=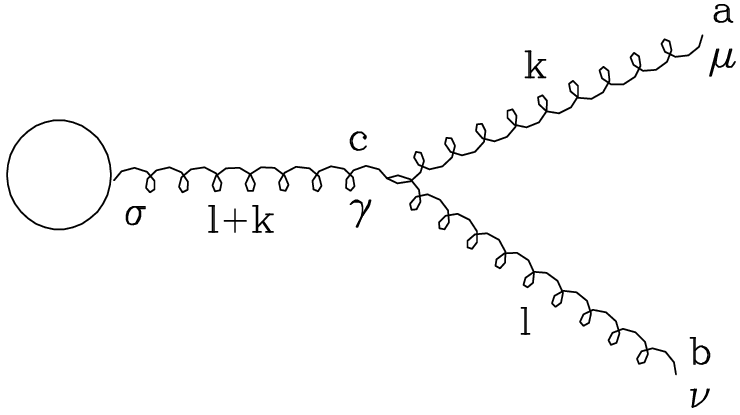,width=0.5\textwidth,clip=}}
\ccaption{}{ \label{ggcoll1}
Gluon splitting}
\end{figure}
In this appendix we will derive the singular part of the square of the
invariant amplitude when two collinear gluons are produced.
In the collinear limit,
the amplitude for the emission of two gluons in the final state can be
decomposed into two parts: the first one contains the graphs where the two
gluons are emitted by a single virtual one (see
Fig.~\ref{ggcoll1}), and the other one contains all the other graphs
\beq
\label{eq:ampcoll}
{\cal A}^{ab} = \ga {\cal A}^{\s}_c(l+k)\, \frac{i
P_{\s\gamma}(k+l)}{(k+l)^2} \, (-g_s)\,
f^{abc} \, \Gamma^{\mu\nu\gamma}(-k,-l,k+l)
+{\cal R}^{\mu\nu}_{ab} \gc \e_{\mu}(k)\, \bar{\e}_{\nu}(l)\;,
\eeq
where $a$ and $b$ are the colour indices of the final gluons,
$P$ is the spin projector of the gluon propagator, $g_s$ is the strong
coupling constant, $f^{abc}$ are
the structure constants of the SU(3) gauge group, $\e$ and $\bar{\e}$
are the polarization vectors of the final gluons, and
$\Gamma^{\mu\nu\gamma}$ is the Lorentz part of the three-gluon vertex
\beq
\label{eq:3gvertex}
\Gamma^{\mu\nu\gamma}(-k,-l,k+l) = (-k+l)^{\gamma}g^{\mu\nu} +
(-2l-k)^{\mu} g^{\nu\gamma} + (2k+l)^{\nu} g^{\mu\gamma}\;.
\eeq
Only the first term of~(\ref{eq:ampcoll}) is singular in the collinear
limit. We want to stress the fact that this term is singular in the
soft limit too. Therefore one has to be careful, when considering the
soft and collinear limit of the square amplitude, not to include this
contribution twice.

We introduce two light-like vectors 
\beqn
t&=&\(\left|\vec{k}+\vec{l}\,\right|,\,\vec{k}+\vec{l}\,\,\) \nonumber \\
\eta&=& c \times\( \frac{1}{\left|\vec{k}+\vec{l}\,\right|}\,,\, -
     \frac{\vec{k}+\vec{l}}{\left|\vec{k}+\vec{l}\,\right|^2}\)
\eeqn
and choose $c=1/4$, so that $2\,t\cdot\eta=1$.
We then decompose
\beq
\label{eq:def_t}
 l^{\mu}+k^{\mu} = t^{\mu} + \xi \, \eta^{\mu}\;,
\eeq
where
\[
 \xi = (l+k)^2 = q^2 y\;.
\]
We will work in the light-cone gauge, characterized by the light-like
vector $\eta$, because, in this gauge (as we will see), the
interference of the divergent term of~(\ref{eq:ampcoll}) and of the finite
term ${\cal R}$ does not contribute to the singular part.

The gluon spin projector is then written
\beq
\label{eq:projector}
  P^{\s\gamma}(p) = -g^{\s\gamma} + \frac{\eta^{\s}p^{\gamma} +
  \eta^{\gamma}p^{\s}}{\eta\cdot p}  \,.
\eeq
We write $l$ and $k$ as 
\beq
\label{eq:def_tperp}
\eqalign{
  k^{\mu} = v\, t^{\mu} + \xi'\eta^{\mu} + \kperp^{\mu}
  \cr
  l^{\mu} = (1-v)\,t^{\mu} + \xi''\eta^{\mu} -\kperp^{\mu}\;,
}
\eeq
with $\kperp$ such that $t\cdot\kperp=\eta\cdot\kperp=0$.
By imposing that $k^2=l^2=0$ and that $(l+k)^2=q^2 y$  we have
\[
\kperp^2 = -v(1-v)\,q^2 y  \hspace{1cm}
\xi' = (1-v)\, q^2 y \hspace{1cm}  \xi'' = v\, q^2 y\;.
\]
From~(\ref{eq:def_t}) and~(\ref{eq:def_tperp}) we
finally obtain
\beq
\label{eq:mom_coll}
\eqalign{
  k^{\mu}=\frac{1}{1-v} \left[ v\,l^{\mu} +(1-2\,v) q^2 y\,\eta^{\mu}
   +\kperp^{\mu} \right]
  \cr
  l^{\mu} = \frac{1}{v}\left[(1-v)\,k^{\mu}-(1-2\,v) q^2 y\,\eta^{\mu}
   -\kperp^{\mu} \right]\;.
}
\eeq
Considering that
\[
  (k+l)^{\gamma} P_{\s\gamma} = 2\,q^2 y\, \eta_{\s} \hspace{1cm}
  k_{\mu} \e^{\mu}(k) = 0  \hspace{1cm}   l_{\nu} \bar{\e}^{\nu}(l) = 0\;,
\]
with the help of eq.~(\ref{eq:mom_coll}) we can rewrite the
amplitude~(\ref{eq:ampcoll})
\dl{
{\cal A}^{ab} = \ga {\cal A}^{\s}_c(l+k)\, \frac{i
P_{\s\gamma}(k+l)}{q^2 y} \, (-g_s)\,
f^{abc}  \right. \nl
{} \times \left.\left[
-2\,\kperp^{\gamma}\, g^{\mu\nu} + \frac{2}{v} \,\kperp^{\mu}\,g^{\nu\gamma} +
\frac{2}{1-v}\,\kperp^{\nu}\, g^{\mu\gamma} +{\cal O}(y)
\right]
+{\cal R}^{\mu\nu}_{ab} \gc \e_{\mu}(k)\, \bar{\e}_{\nu}(l)\;.
}
Observe that the first term is of order $1/\sqrt{y}$ , so that a
singularity with strength $1/y$ can arise only from the square of the
first term, and the interference term does not contribute.
Furthermore, we can now substitute
\[
\eqalign{
  {\cal A}^{\s}_{c}(l+k) \to  {\cal A}^{\s}_{c}(t) \,\equiv\,
  \mbox{\rm tree-level\ amplitude}
  \cr
  P^{\s\gamma}(k+l) \to P^{\s\gamma}(t) =
   -g^{\s\gamma} + \frac{\eta^{\s}t^{\gamma} +
  \eta^{\gamma}t^{\s}}{\eta\cdot t} \equiv -\gperp^{\s\gamma}\;.
}
\]
Remembering that $\eta\cdot\e = \eta\cdot\bar{\e} = 0$, we can write
${\cal A}^{ab}$ in the form
\beq
{\cal A}^{ab} =  {\cal A}_{c\s}(t) \,\frac{g_2}{q^2 y} \, i
f^{abc} \left[
-2\,\kperp^{\s}\, \gperp^{\mu\nu} + \frac{2}{v} \,\kperp^{\mu}\,
\gperp^{\nu\s} +
\frac{2}{1-v}\,\kperp^{\nu}\, \gperp^{\mu\s} \right]
\e_{\mu}(k)\, \bar{\e}_{\nu}(l)\;,
\eeq
where only the term contributing to the singularity has been kept.

By squaring the amplitude and summing over the colours and spins
of the final
gluons, we obtain, for the collinear
singular part
\dl{
{\cal M}_{\rm gg}^{\rm col} = \frac{g_s^2}{q^2}\,\frac{4C_A}{y} \ga -\left[
-2+\frac{1}{v}
+\frac{1}{1-v}+v\,(1-v) \right] g_{\s\s'}\right.\hfill
}
\beq
\label{eq:ggcoll}
\hfill \left. {}- 2\, v\, (1-v) (1-\e)
\left[\frac{k_{\perp\s}k_{\perp\s'}}{\kperp^2}
-\frac{g_{\perp\s\s'}}{2-2\e}\right] \gc  {\cal A}_{c}^{\s}(t)
 {\cal A}_{c}^{*\s'}(t)\;,
\eeq
where we have used the gauge invariance $ t_{\s}\,{\cal
A}_{c}^{\s}(t)=0$ to write the following identity
\[
 {\cal A}_{c}^{\s}(t) {\cal A}_{c}^{*\s'}(t) \,g_{\perp\s\s'}=
   {\cal A}_{c}^{\s}(t) {\cal A}_{c}^{*\s'}(t)\, g_{\s\s'}\;.
\]
The first term of~(\ref{eq:ggcoll}) is recognized to be
the Altarelli-Parisi splitting function for the gluon-gluon process,
in $d=4-2\e$ dimensions. The second term vanishes after azimuthal
average in $4-2\e$ dimensions.

Coming now to our problem, we can further specify the structure of
${\cal A}_{c}^{\s}(t) {\cal A}_{c}^{*\s'}(t)$. 
In fact, by using eq.~(\ref{eq:Mcalssp}), we can write~(\ref{eq:ggcoll}) 
in the following form
\dl{
{\cal M}_{\rm gg}^{\rm col} = g_s^2 \mu^{2\e} \,
\frac{4C_A}{q^2\,y}\, \ga -\left[- 2+\frac{1}{v}
+\frac{1}{1-v}+v\,(1-v) \right] g_{\s\s'}\right.\hfill
}
\beq
\label{eq:ggcollfin}
\hfill \left. {}- 2\, v\, (1-v) (1-\e)
\left[\frac{k_{\perp\s}k_{\perp\s'}}{\kperp^2}
-\frac{g_{\perp\s\s'}}{2-2\e}\right] \gc 
\times {\cal M}_b^{\s\s'} \;.
\eeq

\section{Collinear limit for $g\to q \bar{q}$ splitting}
Following the same steps as in the previous appendix, we can give the
approximation of the square of the amplitude in the limit of a
collinear couple of massless quark-antiquark.
The invariant amplitude is
\dl{
{\cal A} = {\cal A}_{c}^{\s}(k+l)\, \frac{i P_{\rho\s}(k+l)\,\delta^{cc'}}
{(k+l)^2}\,
\bar{u}(k)\(-ig_s\gamma^{\rho}t^{c'}\) v(l) \;, \hfill
}
where $P$ is given by~(\ref{eq:projector}) and $t^c$ are the generators
of SU(3) gauge symmetry.
By squaring this amplitude and summing over the spins and colours
of the final quarks, we obtain
\dl{
{\cal M}_{\rm q\bar{q}}^{\rm col} = \frac{T_F\, 
g_s^2}{q^4 y^2} {\cal A}_{c}^{\s}(k+l) {\cal
A}_{c}^{*\s'}(k+l) P_{\rho\s}(k+l)  P_{\rho'\s'}(k+l)
\tr\(\ds{k}\gamma^{\rho}\ds{l}\gamma^{\rho'}\) \;, \hfill
}
where $t^c$ are normalized such that $\tr\(t^a t^b\) = T_F \delta^{ab}$.

Considering now eqs.~(\ref{eq:def_tperp}), we see that, in the collinear
limit, the trace is of the order of $y$, so that the singular part can be
obtained by putting $y=0$ in the rest of the numerator
\dl{
{\cal M}_{\rm q\bar{q}}^{\rm col} = 
\frac{T_F\,g_s^2}{q^4 y^2} {\cal A}_{c}^{\s}(t) {\cal
A}_{c}^{*\s'}(t)\, g_{\perp\rho\s}  g_{\perp\rho'\s'}
\tr\(\ds{k}\gamma^{\rho}\ds{l}\gamma^{\rho'}\) \;, \hfill
}
where we have used the definition of $t$ given in eq.~(\ref{eq:def_t}).
Evaluating the trace and keeping in the numerator only the terms
proportional to $y$, we obtain
\dl{
{\cal M}_{\rm q\bar{q}}^{\rm col} = \frac{T_F\,g_s^2}{q^4 y^2}\,
 4 \left[ -2 k_{\perp\s} k_{\perp\s'} -\frac{q^2 y}{2} g_{\s\s'}
 \right] {\cal A}_{c}^{\s}(t) {\cal A}_{c}^{*\s'}(t) \;, \hfill
}
that is
\dl{
{\cal M}_{\rm q\bar{q}}^{\rm col} =  \frac{g_s^2}{q^2} \, \frac{4\, T_F}{y}
\ga -\frac{1}{2-2\e}\left[v^2+(1-v)^2-\e\right]g_{\s\s'} \right.  \hfill
}
\beq
\label{eq:qqcoll}
\hfill \left. {} + 2\,v\,(1-v)
\left[\frac{k_{\perp\s}k_{\perp\s'}}{\kperp^2}
-\frac{g_{\perp\s\s'}}{2-2\e}\right] 
\gc
 {\cal A}_{c}^{\s}(t) {\cal A}_{c}^{*\s'}(t) \;. \hfill
\eeq
Here again we can recognize the Altarelli-Parisi kernel for 
$g\,\to\,q \bar{q}$ splitting.

As done before for eq.~(\ref{eq:ggcoll}) , we can specify this formula to
 the problem we are studying. 
With the same substitutions made to go from eq.~(\ref{eq:ggcoll})
to eq.~(\ref{eq:ggcollfin}), we can write
\dl{
{\cal M}_{\rm q\bar{q}}^{\rm col} =  g_s^2 \mu^{2\e}  \,
\frac{4\, T_F}{q^2\,y}
\ga -\frac{1}{2-2\e}\left[v^2+(1-v)^2-\e\right]g_{\s\s'}  \right.  \hfill
}
\beq
\label{eq:qqcollfin}
\hfill \left. {} + 2\,v\,(1-v)
\left[\frac{k_{\perp\s}k_{\perp\s'}}{\kperp^2}
-\frac{g_{\perp\s\s'}}{2-2\e}\right] 
\gc
\times {\cal M}_b^{\s\s'} \;. \hfill
\eeq

\section{Soft limit for the  invariant amplitude 
  $Q\overline{Q} g g$}
\marksection\label{appendix:soft}
In this appendix we will derive the divergent part of the invariant
amplitude for the process
\beq
    Z/\gamma(q) \;\to\; Q(p) + \overline{Q}(p') + g(k) + g(l)
\eeq
in the limit when the momentum $l$ of the gluon is soft.
A soft singularity appears only if the soft gluon is emitted from one of
the external legs.
If the emitting external particle is the gluon, the amplitude of the 
process, in the Feynman gauge, is 
\dl{
{\cal A}^{ab{\rm (g)}}_{ij} 
= {\cal A}^{c\s}_{ij}(l+k)\, \frac{-i}{(k+l)^2} \, (-g_s)\,
f^{abc} \, \Gamma^{\mu\nu}_{\s}(-k,-l,k+l)
\, \e_{\mu}(k)\, \bar{\e}_{\nu}(l)\;,
}
where we have added to 
eq.~(\ref{eq:ampcoll}) the colour indices $i,j$ of the produced quarks. \\
As $l$ goes to zero, this term develops a singularity. By using the
gauge condition $k^{\s}\,{\cal A}_{c\s}^{ij}(k) =0$ and the transversality
$k^{\mu} \e_{\mu}(k)=0$, we can write the amplitude as
\beq
\label{eq:ggsoft}
{\cal A}^{ab{\rm (g)}}_{ij}= g_s f^{abc}\, \frac{k^{\nu}}{k\cdot l}\,
{\cal A}^{c\s}_{ij}(k)\, \e_{\s}(k) \,\bar{\e}_{\nu}(l) + \mbox{\rm
non-singular terms.}
\eeq

Similarly, if we  consider the emission of a soft gluon of colour index
$b$ from an external
quark leg with momentum $p$ and colour index $i$, that is
\[
     Q_n(p+l)\, \to \, Q_i(p) + g_b(l)\,,
\]
we can write the invariant amplitude
\[
{\cal A}_{ij}^{ab{\rm(Q)}} = \bar{u}(p) (-ig_s\gamma^{\nu} t^{b}_{in})
\,\frac{i}{\ds{p}+\ds{l}-m}\, \tilde{\cal {A}}_{nj}^{a \mu}(p+l)
\, \e_{\mu}(k) \,\bar{\e}_{\nu}(l)\,,
\]
where $\tilde{\cal {A}}$ refers to the rest
of the process from which the quark external line takes origin.

In the limit of $l$ going to zero, we can rewrite this amplitude as
\beq
\label{eq:qgsoff}
{\cal A}_{ij}^{ab{\rm(Q)}} = g_s
\,\frac{p^{\nu}}{p\cdot l}\,t^{b}_{in} \, {\cal {A}}_{nj}^{a \mu}(p)
\, \e_{\mu}(k) \,\bar{\e}_{\nu}(l) + \mbox{\rm
non-singular terms,}
\eeq
where we have defined ${\cal {A}}_{nj}^{a \mu}(p) = \bar{u}(p) \tilde{\cal
{A}}_{nj}^{a \mu}(p) $.

In the same way, we can obtain the limit of the amplitude for the
soft emission from an antiquark with momentum $p'$ and colour index $j$
\beq
\label{eq:qbargsoff}
{\cal A}_{ij}^{ab{\rm(\overline{Q})}} = -g_s
\,\frac{p'^{\nu}}{p'\cdot l} \, {\cal {A}}_{in}^{a \mu}(p')\, t^{b}_{nj}
\, \e_{\mu}(k) \,\bar{\e}_{\nu}(l) + \mbox{\rm
non-singular terms.}
\eeq

Considering that ${\cal {A}}_{ij}^{c \s} = t^c_{ij} {\cal {A}}^{\s}$, where 
$ {\cal {A}}^{\s}$ does not contain any colour element, and the
similar ones for eqs.~(\ref{eq:qgsoff}) and~(\ref{eq:qbargsoff}),
we can sum the three amplitudes to obtain
\dl{
{\cal A}_{ij}^{ab} = g_s \ga
 i f^{abc}\, \frac{k^{\nu}}{k\cdot l}\,t^c_{ij}
+ \frac{p^{\nu}}{p\cdot l}\,t^{b}_{in} \, t^{a}_{nj} -
\frac{p'^{\nu}}{p'\cdot l} \,t^{a}_{in}\,t^{b}_{nj} \gc
{\cal {A}}^{\mu}
\, \e_{\mu}(k) \,\bar{\e}_{\nu}(l)  \hfill
}
where we have disregarded the non-singular terms.

By squaring the amplitude and summing over the spins and colours of
 the final gluons and quarks, we have
\dl{
{\cal M}_{\rm gg}^{\rm soft}(l) = g_s^2\,\mu^{2\e}\ga
- C_A
\left[\frac{p\cdot k} {(p\cdot l) \, (k\cdot l)} +
\frac{p'\cdot k} {(p'\cdot l) \, (k\cdot l)} \right]
+ {}\right.  \hfill
}
\beq
\label{eq:lsoft}
\left.{}-2 \(C_F-\frac{C_A}{2}\) \frac{p\cdot p'}{(p\cdot l)\,(p'\cdot l)} +
C_F \left[ \frac{m^2}{(p\cdot l)^2} + \frac{m^2}{(p'\cdot l)^2} \right]
 \gc
\times {\cal M}^\s_\s
\eeq
where we have made use of eq.~({\ref{eq:Mcalssp}).

The same result applies in the case of $k$ soft, once the interchange $l
\leftrightarrow k $ is made.

\section{One-loop scalar integrals}
\marksection\label{sec:virtual}
We can classify
the different types of scalar integrals according to the number of massive 
propagators in the loop and according to the ``shape'' of the loop: boxes
(B) and triangles (T).
We introduce the following kinematical invariants
\beq
\eqalign{
\label{eq:def_sig1sig2}
 \su = (q-p')^2 -m^2 = q^2(1-x_2)  \cr
 \sd = (q-p)^2 -m^2 = q^2(1-x_1)\cr
 \st = (q-k)^2 = q^2(1-x_g)\;,
}
\eeq
where $x_1$ and $x_2$ are defined by~(\ref{eq:x1x2y}) and
$x_g$ by~(\ref{eq:xg_def}), and
\dl{ 
\lpm = \frac{1}{2}\( 1 \pm \sqrt{1-\frac{4m^2}{q^2}}\) \equiv
          \frac{1}{2}\( 1 \pm \d \)  
\hfill}
\dl{
\zpm = \frac{1}{2}\( 1 \pm \sqrt{1-\frac{4m^2}{\st}}\)  \equiv
           \frac{1}{2}\( 1\pm \dpr \) 
\hfill}
\dl{
 \ropm = \frac{1}{2}\left[ \a_{1} \pm
 \sqrt{\a_{1}^2-\frac{4m^2}{q^2}}\right]  \hspace{1cm} {\rm with:\ \ }
     \a_{1}=1-\frac{\su}{q^2}
\hfill}
\dl{
  \eta_\pm = \frac{1}{2}\(\dpr \pm \d \)  \,.
\hfill}
Here we also give the absorptive parts of the integrals,
although they do not contribute to the cross section.
The integrals are computed in $d=4-2\e$ dimensions. Terms
of order $\e$ or higher have been dropped.

Defining the dilogarithm function as
\[                    \li{x} = -\int_0^x dz \, \frac{\log(1-z)}{z}
\]
and collecting the same factor
\[
N(\e)=\frac{i}{16\,\pi^2}\ (4\pi)^\e \ \Gamma(1+\e) = i\, N_\e
\]
in front of each expression, we obtain
\beqn
B_{2m} &\equiv& \int \frac{d^dl}{(2\pi)^d}\ \frac{1}{l^2} \ \frac{1}{(l-k)^2} \
\frac{1}{(l+p-q)^2-m^2} \ \frac{1}{(l+p)^2-m^2} \nonumber \\
& = & N(\e)\, \(m^2\)^{-\e}
\frac{1}{\su \sd}  \left\{ \frac{1}{\e^2}+\frac{1}{\e}\(
    \log\frac{m^2}{\su}+\log\frac{m^2}{\sd}\) + 2
  \log\frac{m^2}{\su}\, \log\frac{m^2}{\sd} \right. \nonumber \\
&& \left. {}-\frac{5}{3}\pi^2
  -\log^2 \frac{\lp}{\lm} + 2 \pi i \left[
    \frac{1}{\e}+\log\frac{m^2}{\su}+\log\frac{m^2}{\sd} + \log
    \frac{\lp}{\lm} \right] \right\} 
\\
B_{3m}&\equiv& \int \frac{d^dl}{(2\pi)^d}\ \frac{1}{l^2} \
\frac{1}{(l+p)^2-m^2} \
\frac{1}{(l+p-q)^2-m^2} \ \frac{1}{(l-p')^2-m^2} \nonumber\\
& =& N(\e)\ (m^2)^{-\e} \frac{1}{\sd \st \dpr }
\left\{\frac{1}{\e}\log\frac{\zm}{\zp} +
\(2\log\frac{m^2}{\sd}+\log\frac{m^2}{\st}
\)  \log\frac{\zm}{\zp}   \right.  \nonumber\\
&&
{}-2\li{\zm} -\log^2\zm -2\li{-\frac{\lm}{\eta_{+}}}-2\li{\frac{\eta_{+}}{\lp}}
-\log^2\frac{\lp}{\eta_{+}} \nonumber\\
&&{}+2\li{\frac{\lm}{\eta_{-}}}+2\li{-\frac{\eta_{-}}{\lp}}
+\log^2\(-\frac{\lp}{\eta_{-}}\) +2\li{-\frac{\zm}{\dpr}}
-\frac{\pi^2}{2}  \nonumber\\
&&{}+ 2\log\(-\frac{\eta_{+}}{\eta_{-}}\) \log\frac{\lm}{\lp} +\log^2\zp
-2\log\dpr\log\zm +\log^2\dpr  \nonumber\\
&&\left.{}+ i\pi\left[\frac{1}{\e} +2\log\frac{q^2}{\sd}
 +4\log\eta_{+} -2\log\dpr +2\log\frac{\zm}{\zp} \right]
\right\}
\\
T_{2m}^q &\equiv& \int \frac{d^dl}{(2\pi)^d}\ \frac{1}{l^2} \
\frac{1}{(l-p')^2-m^2} \
\frac{1}{(l+p+k)^2-m^2}  \nonumber \\
& = & N(\e)\ (m^2)^{-\e}
 \frac{-1}{q^2\sqrt{\a_{1}^2-\frac{4m^2}{q^2}}}
\left\{\li{1-\frac{1}{\rop}}+\li{-\frac{\rop}{\lp-\rop}}{} \right.
\nonumber \\
&&{}+ \li{\frac{\lp-\rop}{1-\rop}}-\li{\frac{\rop-\lm}{\rop}}-
\li{\frac{1-\rop}{\lm-\rop}}-\li{\rom}  \nonumber \\
&&{} -\li{\frac{\lp-\rom}{1-\rom}}  -\li{-\frac{\rom}{\lp-\rom}}+
\li{\frac{\lm}{\lm-\rom}}\nonumber \\
&&{} + \li{\frac{\rom-\lm}{1-\lm}}+
\frac{1}{2}\log^2\frac{\lp-\rop}{1-\rop}
-\frac{1}{2}\log^2\frac{\rop-\lm}{\rop}-\frac{1}{2}\log^2\rom\nonumber \\
&&{}-\log\rom\log\frac{1-\rom}{\rom}
-\frac{1}{2}\log^2\frac{\lp-\rom}{1-\rom}+\frac{1}{2}\log^2
\frac{\rom-\lm}{1-\lm}\nonumber \\
&&{}-\log\frac{\rom-1}{\lm-\rom}\log\frac{\lm-1}{\lm-\rom}
+\log\frac{-\rom}{\lm-\rom}\log\frac{-\lm}{\lm-\rom} + \frac{\pi^2}{6}
\nonumber \\
&&\left. {}+i \pi\left[
2\,\log\frac{\lp-\rom}{\lp-\rop} +\log\frac{1-\rop}{1-\rom}
\right] \right\}
\\
T_{2m}^{q-k} &\equiv& \int \frac{d^dl}{(2\pi)^d}\ \frac{1}{l^2} \
\frac{1}{(l-p')^2-m^2} \
\frac{1}{(l+p)^2-m^2}  \nonumber \\
& = &N(\e)\,\(m^2\)^{-\e}\frac{1}{\st \dpr} \left\{
\frac{1}{\e}\log\frac{\cm}{\cp} +\log\frac{m^2}{\st}\log\frac{\cm}{\cp}-
\frac{1}{2}\log^2\cm \right.\nonumber \\
&&{} +\frac{1}{2}\log^2 \cp -\log\dpr\log\frac{\cm}{\cp}
+\li{-\frac{\cm}{\dpr}}+\li{\frac{\dpr}{\cp}}
+\frac{1}{2}\log^2\frac{\cp}{\dpr}\nonumber \\
&&\left.{}-\frac{5}{6}\pi^2 + i \pi \left[ \frac{1}{\e} +
\log\frac{m^2}{\st}-2\log\dpr \right] \right\}
\\
T_{2m} &\equiv& \int \frac{d^dl}{(2\pi)^d}\ \frac{1}{l^2} \
\frac{1}{(p+l)^2-m^2} \
\frac{1}{(p+k+l)^2-m^2}  \nonumber \\
&=& N(\e)\,\(m^2\)^{-\e}\frac{1}{\su}\left\{
\li{-\frac{\su}{m^2}} +\log\frac{\su}{m^2} \log\(1+\frac{\su}{m^2}
\)\right. \phantom{qqqqqqqqq}\nonumber \\
&&\left.{}-i \pi \log\(1+\frac{\su}{m^2}\)\right\}
\\
T_{1m} &\equiv& \int \frac{d^dl}{(2\pi)^d}\ \frac{1}{l^2} \
\frac{1}{(l+k)^2} \
\frac{1}{(l+p+k)^2-m^2}  \nonumber \\
 &=& N(\e)\,\(m^2\)^{-\e} \frac{1}{\su} \left\{
\frac{1}{2\e^2}+\frac{1}{\e}\log\frac{m^2}{\su} +
\log\frac{m^2}{\su}\log\(1+\frac{m^2}{\su}\) {}\right.\nonumber \\
&&\left.{}-\li{-\frac{\su}{m^2}}-\frac{5}{6}\pi^2 + i\pi \left[ \frac{1}{\e}
+\log\(1+\frac{\su}{m^2}\) + 2\log\frac{m^2}{\su}\right]
\right\}
\\
T_{3m} &\equiv& \int \frac{d^dl}{(2\pi)^d}\ \frac{1}{(l-p')^2-m^2} \
\frac{1}{(l+p-q)^2-m^2} \
\frac{1}{(l+p)^2-m^2}  \nonumber \\
& =& N(\e) \(m^2\)^{-\e} \frac{1}{2}\, \frac{1}{\st-q^2}
\left\{ \log^2\(\frac{1}{\cm}-1\) -
\log^2\(\frac{1}{\lm}-1\)   {}\right.\nonumber \\
&&\left.{}-2 i \pi \left[\log\(\frac{1}{\cm}-1\)-
\log\(\frac{1}{\lm}-1\) \right] \right\}
\eeqn

A partial check of the correctness of the above formulae can be
performed in the following way.
We consider first a check of $B_{2m}$. To this purpose, define $I$ to
be
\dl{
\!I\equiv\!\int\!\frac{d^dl}{(2\pi)^d}\ \frac{1 + A(l-k)^2 + B \left[
(l+p-q)^2-m^2 \right] +C\left[ (l+p)^2-m^2\right]}{
 l^2 (l-k)^2 \left[(l+p-q)^2-m^2\right] \left[(l+p)^2-m^2\right]} =
\hfill}
\dl{
\!\phantom{I} =\!\int\! \frac{d^dl}{(2\pi)^d}\  \frac{
1 + B [q^2-2p\cdot q] + 2 l\cdot \left[ -A k + B (p-q) + Cp\right]
+l^2(A+B+C)
}{ l^2 (l-k)^2 \left[(l+p-q)^2-m^2\right] \left[(l+p)^2-m^2\right]}
\hfill}
If we impose that $I$ has no infrared and collinear divergences, then
$$
\left\{
 \eqalign{
  1 + B [q^2-2p\cdot q] = 0 \cr
  k \cdot  \left[ -A k + B (p-q) + Cp\right] = 0
 }
\right.
$$
Solving this system
$$
\left\{
 \eqalign{
   B = -\frac{1}{\sd} \cr
   C = -\frac{1}{\su}\;.
 }
\right.
$$
So $I$ can be rewritten as
\dl{
I =
 B_{2m} + A T_{2m}^{'q} -\frac{1}{\sd} T_{1m} -\frac{1}{\su}
T_{1m}^{'}
\hfill}
where the primed quantities are the same as before, with the
substitution $p\leftrightarrow p'$, that is $\su \leftrightarrow \sd$.
The integral $I$ is now convergent and the cancellation of the divergent
part of the right-hand side
 can be checked directly (both in the real part and in the
absorptive one).
As far as the finite terms are concerned, the integral $I$ can be
reduced to a one-variable integral, using Feynman parametrization,
and then integrated numerically to check the identity.

With the same reasoning, we can check $B_{3m}$. We introduce the
integral
\dl{
\!I\equiv\!\int\! \frac{d^dl}{(2\pi)^d}\ \frac{1 + A \left[(l+p)^2-m^2\right]
+ B \left[(l+p-q)^2-m^2\right] + C \left[(l-p')^2-m^2\right] }{
 l^2 \left[(l+p)^2-m^2\right] \left[(l+p-q)^2-m^2\right]
 \left[(l-p')^2-m^2\right]  }
\hfill}
\dl{
\!\phantom{I} =\!\int\! \frac{d^dl}{(2\pi)^d}\  \frac{
1 + B \left[q^2-2p\cdot q\right] + 2 l\cdot \left[ A p + B (p-q) - Cp'\right]
+l^2(A+B+C)
}{ l^2 \left[(l+p)^2-m^2\right] \left[(l+p-q)^2-m^2\right]
 \left[(l-p')^2-m^2\right]  }\;.
\hfill}
This integral has only soft divergences, which can be removed if we
require that
\[
   1+ B\left[q^2-2p\cdot q\right] = 0   \Longrightarrow  B = -\frac{1}{\sd}\;.
\]
Thus $I$ becomes
\dl{
I  = B_{3m} + A T_{2m}^{'} -\frac{1}{\sd} T_{2m}^{q-k} +C
T_{2m}^{'q} \;.
\hfill}
The rest of the check is the same as before.

\section{List of integrals for the soft contributions}
\marksection\label{appendix:soft_integrals}
We now summarize the values of the integrals required to
isolate the singular pieces of the four-jet cross section, in the soft-gluon
limit
\dl{
I_1(x,h) =
\int_0^x d y \int_0^1 dv\,[v(1-v)]^{-\e} y^{-\e}  \frac{1} {y\,[y+h\,v]}=
\hfill\cr
\phantom{I_1(x,h) \quad} =
  \frac{1}{2h} \ga
\frac{1}{\e^2}-\frac{1}{\e} \log h - \log^2\frac{x}{h}
+\frac{1}{2}\log^2 h -\frac{\pi^2}{2}-2 \li{-\frac{x}{h}} \gc
 + {\cal O}(\e)\hfill
}
\dl{
I_2(x,h) =
\int_0^x d y \int_0^1  dv\,[v(1-v)]^{-\e} y^{-\e}  \frac{1} {[y+h\, v]^2}=
\hfill\cr
\phantom{I_2(x,h) \quad} =
 \frac{1}{2h} \ga -\frac{1}{\e} -2\log\(1+\frac{h}{x}\) +\log h \gc
 + {\cal O}(\e)\hfill
}
\dl{
I_3 =    \frac{1}{N_{\thp}}
\int_0^\pi d\thp\,(\sin\thp)^{-2\e} 
\int_0^x d y \int_0^1
  dv\,[v(1-v)]^{-\e} y^{-\e}  {}\nl
\phantom{\quad I_3 = }{} \times \frac{1} {y + h\, v} \,
\frac{1}{y-c\cos\thp\,\sqrt{y}\,\sqrt{v} + g\, v} = \hfill
\cr
\phantom{\quad I_3} =    \frac{1}{N_{\thp}} \ga
 -\frac{1}{2\e} I_{\e}
+ I_{\thp} -\frac{1}{2} \left[-I_x +I_{1/x} \right] \gc  + {\cal O}(\e)
\hfill
}
where $I_{\e},I_\thp,I_x$ and $I_{1/x}$ are finite quantities, defined
by
\dl{
I_{\e} = \frac{\pi}{K} \frac{\(1+\dpr^2\)\(1-\dpr^2\)}{4\dpr}
\(1-x_g\) \log\(\frac{\zp}{\zm}\)^2    \hfill
}
\dl{
I_{\thp} = \int_0^{\frac{\pi}{2}} d\thp\, \log\sin\thp \
\frac{2}{(g-h)^2+h\(c\cos\thp\)^2} {}\hfill\cr
\hfill{}\times\ga
\frac{2\,c\,(g+h)\cos\thp} {\sqrt{4g-c^2\cos^2\thp}}
\arctan\frac{c\cos\thp}{\sqrt{4g-c^2\cos^2\thp}} + (g-h)\log\frac{g}{h}
\gc}
\dl{ I_x = \int_{0}^{x} dt\,
 \frac{\log t}{\(h+t\)}  \frac{\pi}{\sqrt{\(t+g\)^2-c^2t}}
\hfill}
\dl{
I_{1/x}=  \int_0^{\frac{1}{x}} dt\,
 \frac{-2\log x -\log t} {\(1+ht\)}
 \frac{\pi}{\sqrt{\(1+gt\)^2-c^2t}} \,.
\hfill}
For the definition of the constants appearing in these integrals, see
Section~\ref{sec:sc_limit} and Appendix~\ref{sec:virtual}.

\section{Results}
We implemented our analytical result in a FORTRAN program, which
behaves like a ``partonic'' Monte Carlo generator, analogous to
the program EVENT \cite{YellowBook}.
We collect here some results obtained with our code, with which
future users of the program may, eventually, compare their results.
Furthermore, since for this kind of calculations it would be difficult
to perform analytical comparisons, the only possible alternative
is to choose a few shape variables, and compare numerical results,
in the spirit of what has been done in ref.~\cite{LEP2} for the case of
the massless calculation.

We include in these results only the contributions from cut graphs
of A-type, that is to say, from cut graphs in which the weak current couples
to the same heavy-flavour loop, and there is a single $Q\overline{Q}$ pair
in the final state, which is the really hard part of the calculation.
For the contributions involving two heavy-quark pairs in the final
state, it is easier to compare directly the value of the matrix
elements squared (this part of our program was in fact checked in this way
with the program of ref.~\cite{Ballestrero}).

We have chosen a set of shape variables
for which it should be easy to obtain quite accurate numerical results.
We have fixed the centre-of-mass energy to be 100 GeV, and the mass of the
heavy quark has been taken to be equal to 1, 10, 20 and 30 GeV.
We present separately the results for a hypothetical vector boson
with purely axial or purely vector couplings,
normalized to the massless total cross section at the zeroth order in $\as$.
We have chosen the following shape variables: the thrust
$t$, the $c$ parameter, the mass of the heavy jet squared $M_h^2$
(according to the thrust axis), the energy--energy correlation EEC, the
three-jet fractions according to the E, EM, JADE, and DURHAM schemes.
For $t$, $c$, $M_h^2$ and EEC we present moments, instead of distributions,
because they can be obtained with higher precision.
For thrust, for example, we thus compute, according to the notation of
Section~\ref{sec:outlinecalc}
\beq
\int dT_{V/A} \, (1-t)^n = \left(\frac{\as}{2\pi}\right)   A^t_{V/A}(n) +
                        \left(\frac{\as}{2\pi}\right)^2 B^t_{V/A}(n)\;.
\eeq
We will further decompose
\beq
B^t_{V/A} = B_{V/A,\,C_A}^{t} + B_{V/A,\,C_F}^{t} +B_{V/A,\,T_F}^{t}\;,
\eeq
where the $C_A$, $C_F$ and $T_F$ subscripts denote the $C_F C_A$,
$C_F^2$ and $n_{\rm f} C_F T_F$ colour components.
For some shape variables, the presence of massive particles in the
final state may introduce ambiguities in the definition, owing
to the fact that, in the massless case, energy and momentum can be 
interchanged.
We thus refer to the exact definitions given in ref.~\cite{YellowBook}
for $t$, $c$,  $M_h^2$ and in ref.~\cite{LEP2} for the EEC.
Moments are defined as
\beqn
\int dT_{V/A}\; c^n &=& \left(\frac{\as}{2\pi}\right)   A^c_{V/A}(n) +
                        \left(\frac{\as}{2\pi}\right)^2 B^c_{V/A}(n)\;,
\nonumber \\
\int dT_{V/A} \left(\frac{M_h^2-m^2}{q^2}\right)^n &=&
\left(\frac{\as}{2\pi}\right)   A^{M_h}_{V/A}(n) +
                        \left(\frac{\as}{2\pi}\right)^2 B^{M_h}_{V/A}(n)\;,
\nonumber \\
\int dT_{V/A} \sum_{ij} \frac{E_i E_j}{q^2} \cos^k\theta_{ij}
 \sin^{2+n}\theta_{ij} &=&
             \left(\frac{\as}{2\pi}\right)   A^{\rm EEC}_{V/A}(n,k) +
                   \left(\frac{\as}{2\pi}\right)^2 B^{\rm EEC}_{V/A}(n,k)\;.
\nonumber
\eeqn
where the sum runs over all the final particles.

Clusters are defined in the following way. There is a resolution parameter $y$,
which is computed for every pair of particles in the final state.
One finds the pair for which $y$ is minimum. 
If $y < y_{cut}$ the two particles are combined into a single
pseudo-particle by adding up their four momenta. One thus computes
\beq
\int dT_{V/A} \delta_{N_{\rm X}(y_{cut}),\,3}=
             \left(\frac{\as}{2\pi}\right)   A^{\rm X}_{V/A}(y_{cut}) +
                   \left(\frac{\as}{2\pi}\right)^2 B^{\rm X}_{V/A}(y_{cut})\;,
\eeq
where ${\rm X}$ stands for E, EM, JADE or DURHAM, and 
$N_{\rm X}(y_{cut})$ is the number of pseudo-particles in the final state
after the clustering procedure.
The various clustering algorithms differ by the definition of the 
resolution parameter $y$
\beqn
\mbox{E}&:& \frac{(p_i+p_j)^2}{q^2}\;,
\nonumber \\
\mbox{EM}&:& 2\; \frac{p_i\cdot p_j}{q^2}\;,
\nonumber \\
\mbox{JADE}&:& 2\; \frac{E_i E_j}{q^2} (1-\cos\theta_{ij})\;,
\nonumber \\
\mbox{DURHAM}&:& 2\min{\(E_i^2,E_j^2\)}\frac{1}{q^2}
 (1-\cos\theta_{ij})\;.
\eeqn
Observe that the E scheme is not infrared-safe if $y_{cut}<m^2/q^2$.
In fact, in this case, the configuration made up of two heavy quarks
plus a soft
gluon cannot be reduced to two pseudo-particles, since the recombination
parameter will fail the cut, for any pair containing a massive quark.
The cancellation of soft divergences cannot therefore work for these values
of the cut parameter.

We have chosen the renormalization scale $\mu=E$, and $n_f=5$.
The results are given in Tables \ref{th} to \ref{DU}.
\clearpage
 \begin{table}[htbp]
 \begin{center}
 \footnotesize
\begin{tabular}{||c|c|c|c|c||}
 \hline\hline
 $n$&$m/E=0.01$ & $m/E=0.1$ & $m/E=0.2$                     &$m/E=0.3$\\ \hline
 &\multicolumn{4}{c||}{$\phantom{\Bigl[}B_{V,\,C_A  }^{t            }
 (n)$}
 \\ \hline
   1&
 $71.53 \pm 0.045$ & $57.14 \pm 0.024$ & $39.17 \pm 0.013$ & $21.83 \pm 0.006$ \\                                                                                          
   2&
 $5.303 \pm 0.005$ & $4.47 \pm 0.0034$ & $3.026 \pm 0.002$ & $1.524 \pm 8\,10^{-4}$ \\                                                                                     
   3&
 $0.8056 \pm 0.0012$ & $0.6887 \pm 8\,10^{-4}$ & $0.4582 \pm 5\,10^{-4}$ & $0.2127 \pm 2\,10^{-4}$ \\                                                                      
   4&
 $0.1604 \pm 3.4\,10^{-4}$ & $0.1381 \pm 2.4\,10^{-4}$ & $0.09096 \pm 1.4\,10^{-4}$ & $0.04003 \pm 6\,10^{-5}$ \\                                                          
   5&
 $0.03668 \pm 1\,10^{-4}$ & $0.03174 \pm 7\,10^{-5}$ & $0.02079 \pm 4\,10^{-5}$ & $0.008845 \pm 2\,10^{-5}$ \\                                                             
 \hline
 &\multicolumn{4}{c||}{$\phantom{\Bigl[}B_{V,\,C_F  }^{t            }
 (n)$}
 \\
\hline
 \hline
 \end{tabular}
 \caption{The thrust $t$.                                   }
 \label{th}
 \end{center}
 \end{table}
 \begin{table}[htbp]
 \begin{center}
 \footnotesize
 \begin{tabular}{||c|c|c|c|c||}
 \hline\hline
 $n$&$m/E=0.01$ & $m/E=0.1$ & $m/E=0.2$                     &$m/E=0.3$\\ \hline
 &\multicolumn{4}{c||}{$\phantom{\Bigl[}B_{V,\,C_A  }^{M_h          }
 (n)$}
 \\ \hline
   1&
 $68.32 \pm 0.05$ & $56.71 \pm 0.03$ & $38.45 \pm 0.02$ & $19.46 \pm 0.008$ \\                                                                                             
   2&
 $4.591 \pm 0.006$ & $4.014 \pm 0.004$ & $2.642 \pm 0.0035$ & $1.029 \pm 0.0024$ \\                                                                                        
   3&
 $0.6336 \pm 0.0013$ & $0.5549 \pm 0.001$ & $0.3565 \pm 0.0013$ & $0.1072 \pm 0.001$ \\                                                                                    
   4&
 $0.1156 \pm 3.6\,10^{-4}$ & $0.1002 \pm 3\,10^{-4}$ & $0.0641 \pm 5\,10^{-4}$ & $0.0167 \pm 4.5\,10^{-4}$ \\                                                              
   5&
 $0.02447 \pm 1\,10^{-4}$ & $0.02084 \pm 1\,10^{-4}$ & $0.01346 \pm 1.8\,10^{-4}$ & $0.0037 \pm 2\,10^{-4}$ \\                                                             
 \hline
 &\multicolumn{4}{c||}{$\phantom{\Bigl[}B_{V,\,C_F  }^{M_h          }
 (n)$}
 \\
\hline
 \hline
 \end{tabular}
 \caption{The mass of the heavy jet squared $M_h^2$.        }
 \label{mh}
 \end{center}
 \end{table}
 \begin{table}[htbp]
 \begin{center}
 \footnotesize
 \begin{tabular}{||c|c|c|c|c||}
 \hline\hline
 $n$&$m/E=0.01$ & $m/E=0.1$ & $m/E=0.2$                     &$m/E=0.3$\\ \hline
 &\multicolumn{4}{c||}{$\phantom{\Bigl[}B_{V,\,C_A  }^{c            }
 (n)$}
 \\ \hline
   1&
 $303.3 \pm 0.18$ & $233.9 \pm 0.09$ & $155.2 \pm 0.05$ & $84.7 \pm 0.023$ \\                                                                                              
   2&
 $70.97 \pm 0.06$ & $58.25 \pm 0.03$ & $38.33 \pm 0.018$ & $19.18 \pm 0.007$ \\                                                                                            
   3&
 $30.55 \pm 0.03$ & $25.56 \pm 0.02$ & $16.63 \pm 0.01$ & $7.789 \pm 0.004$ \\                                                                                             
   4&
 $16.36 \pm 0.02$ & $13.85 \pm 0.012$ & $8.955 \pm 0.007$ & $4.008 \pm 0.003$ \\                                                                                           
   5&
 $9.784 \pm 0.012$ & $8.341 \pm 0.008$ & $5.379 \pm 0.005$ & $2.335 \pm 0.002$ \\                                                                                          
 \hline
 &\multicolumn{4}{c||}{$\phantom{\Bigl[}B_{V,\,C_F  }^{c            }
 (n)$}
 \\
\hline
 \hline
 \end{tabular}
 \caption{The $c$ parameter.                                }
 \label{cp}
 \end{center}
 \end{table}
 \begin{table}[htbp]
 \begin{center}
 \footnotesize
 \begin{tabular}{||c|c|c|c|c||}
 \hline\hline
 $n$&$m/E=0.01$ & $m/E=0.1$ & $m/E=0.2$                     &$m/E=0.3$\\ \hline
 &\multicolumn{4}{c||}{$\phantom{\Bigl[}B_{V,\,C_A  }^{{\rm EEC}    }
 (n,0)$}
 \\ \hline
   0&
 $202.3 \pm 0.12$ & $154.4 \pm 0.06$ & $97.08 \pm 0.03$ & $46.81 \pm 0.013$ \\                                                                                             
   1&
 $143.1 \pm 0.1$ & $117 \pm 0.05$ & $77.18 \pm 0.026$ & $38.2 \pm 0.01$ \\                                                                                                 
   2&
 $115.9 \pm 0.09$ & $97.55 \pm 0.04$ & $66.33 \pm 0.022$ & $33.55 \pm 0.009$ \\                                                                                            
   3&
 $99.79 \pm 0.08$ & $85.27 \pm 0.04$ & $59.11 \pm 0.02$ & $30.38 \pm 0.008$ \\                                                                                             
   4&
 $88.92 \pm 0.08$ & $76.65 \pm 0.04$ & $53.84 \pm 0.02$ & $28.01 \pm 0.007$ \\                                                                                             
   5&
 $80.95 \pm 0.07$ & $70.18 \pm 0.04$ & $49.76 \pm 0.018$ & $26.13 \pm 0.007$ \\
\hline
 \hline
 \end{tabular}
 \caption{The energy--energy correlation EEC, $k=0$.        }
 \label{e0}
 \end{center}
 \end{table}
 \begin{table}[htbp]
 \begin{center}
 \footnotesize
 \begin{tabular}{||c|c|c|c|c||}
 \hline\hline
 $n$&$m/E=0.01$ & $m/E=0.1$ & $m/E=0.2$                     &$m/E=0.3$\\ \hline
 &\multicolumn{4}{c||}{$\phantom{\Bigl[}B_{V,\,C_A  }^{{\rm EEC}    }
 (n,1)$}
 \\ \hline
   0&
 $-26.06 \pm 0.025$ & $-21.45 \pm 0.014$ & $-14.12 \pm 0.008$ & $-7.006 \pm 0.003$ \\                                                                                      
   1&
 $-9.56 \pm 0.016$ & $-8.967 \pm 0.01$ & $-6.413 \pm 0.006$ & $-3.152 \pm 0.0023$ \\                                                                                       
   2&
 $-4.827 \pm 0.013$ & $-4.892 \pm 0.008$ & $-3.785 \pm 0.0045$ & $-1.898 \pm 0.002$ \\                                                                                     
   3&
 $-2.894 \pm 0.011$ & $-3.086 \pm 0.008$ & $-2.549 \pm 0.004$ & $-1.316 \pm 0.0017$ \\                                                                                     
   4&
 $-1.93 \pm 0.01$ & $-2.133 \pm 0.007$ & $-1.858 \pm 0.0036$ & $-0.9876 \pm 0.0015$ \\                                                                                     
   5&
 $-1.38 \pm 0.01$ & $-1.57 \pm 0.007$ & $-1.427 \pm 0.0034$ & $-0.779 \pm 0.0014$ \\
\hline
 \hline
 \end{tabular}
 \caption{The energy--energy correlation EEC, $k=1$.        }
 \label{e1}
 \end{center}
 \end{table}
 \begin{table}[htbp]
 \begin{center}
 \footnotesize
 \begin{tabular}{||c|c|c|c|c||}
 \hline\hline
 $y_{cut}$&$m/E=0.01$ & $m/E=0.1$ & $m/E=0.2$               &$m/E=0.3$\\ \hline
 &\multicolumn{4}{c||}{$\phantom{\Bigl[}B_{V,\,C_A  }^{{\rm E}      }(y_{cut})$}\\ \hline
0.01&
 $1357 \pm 1.5$ & $-$ & $-$ & $-$ \\                                                                                                                                       
0.05&
 $341.3 \pm 0.5$ & $374 \pm 0.3$ & $817.1 \pm 0.3$ & $-$ \\                                                                                                                
0.10&
 $133.7 \pm 0.3$ & $138.6 \pm 0.17$ & $164.3 \pm 0.1$ & $459.5 \pm 0.17$ \\                                                                                                
0.15&
 $60.75 \pm 0.18$ & $63.1 \pm 0.13$ & $64.29 \pm 0.07$ & $79.06 \pm 0.04$ \\                                                                                               
0.20&
 $27.13 \pm 0.16$ & $29 \pm 0.12$ & $29 \pm 0.05$ & $24.54 \pm 0.027$ \\                                                                                                   
 \hline
 &\multicolumn{4}{c||}{$\phantom{\Bigl[}B_{V,\,C_F  }^{{\rm E}      }(y_{cut})$}\\
\hline
 \hline
 \end{tabular}
 \caption{The E clustering algorithm.                       }
 \label{EA}
 \end{center}
 \end{table}
 \begin{table}[htbp]
 \begin{center}
 \footnotesize
 \begin{tabular}{||c|c|c|c|c||}
 \hline\hline
 $y_{cut}$&$m/E=0.01$ & $m/E=0.1$ & $m/E=0.2$               &$m/E=0.3$\\ \hline
 &\multicolumn{4}{c||}{$\phantom{\Bigl[}B_{V,\,C_A  }^{{\rm EM}     }(y_{cut})$}\\ \hline
0.01&
 $1340 \pm 1.7$ & $1135 \pm 0.8$ & $803.4 \pm 0.4$ & $455.7 \pm 0.18$ \\                                                                                                   
0.05&
 $333.2 \pm 0.5$ & $291.2 \pm 0.4$ & $203.4 \pm 0.18$ & $101.6 \pm 0.07$ \\                                                                                                
0.10&
 $126.3 \pm 0.3$ & $111.8 \pm 0.23$ & $73.76 \pm 0.11$ & $29.24 \pm 0.04$ \\                                                                                               
0.15&
 $54.24 \pm 0.3$ & $48 \pm 0.17$ & $30.47 \pm 0.12$ & $8.176 \pm 0.027$ \\                                                                                                 
0.20&
 $22.02 \pm 0.2$ & $18.85 \pm 0.14$ & $10.77 \pm 0.08$ & $1.52 \pm 0.024$ \\                                                                                               
 \hline
 &\multicolumn{4}{c||}{$\phantom{\Bigl[}B_{V,\,C_F  }^{{\rm EM}     }(y_{cut})$}\\
\hline
 \hline
 \end{tabular}
 \caption{The EM clustering algorithm.                      }
 \label{EM}
 \end{center}
 \end{table}
 \begin{table}[htbp]
 \begin{center}
 \footnotesize
 \begin{tabular}{||c|c|c|c|c||}
 \hline\hline
 $y_{cut}$&$m/E=0.01$ & $m/E=0.1$ & $m/E=0.2$               &$m/E=0.3$\\ \hline
 &\multicolumn{4}{c||}{$\phantom{\Bigl[}B_{V,\,C_A  }^{{\rm JADE}   }(y_{cut})$}\\ \hline
0.01&
 $1352 \pm 2$ & $1088 \pm 1.6$ & $727.2 \pm 0.5$ & $389.2 \pm 0.26$ \\                                                                                                     
0.05&
 $328.6 \pm 0.6$ & $275 \pm 0.5$ & $180 \pm 0.2$ & $82.34 \pm 0.09$ \\                                                                                                     
0.10&
 $124.2 \pm 0.3$ & $106 \pm 0.26$ & $66.06 \pm 0.15$ & $24.91 \pm 0.05$ \\                                                                                                 
0.15&
 $54.1 \pm 0.3$ & $46.84 \pm 0.26$ & $28.77 \pm 0.1$ & $8.675 \pm 0.03$ \\                                                                                                 
0.20&
 $22.86 \pm 0.2$ & $19.84 \pm 0.14$ & $12.34 \pm 0.09$ & $2.98 \pm 0.03$ \\                                                                                                
 \hline
 &\multicolumn{4}{c||}{$\phantom{\Bigl[}B_{V,\,C_F  }^{{\rm JADE}   }(y_{cut})$}\\
\hline
 \hline
 \end{tabular}
 \caption{The JADE clustering algorithm.                    }
 \label{JA}
 \end{center}
 \end{table}
 \begin{table}[htbp]
 \begin{center}
 \footnotesize
 \begin{tabular}{||c|c|c|c|c||}
 \hline\hline
 $y_{cut}$&$m/E=0.01$ & $m/E=0.1$ & $m/E=0.2$               &$m/E=0.3$\\ \hline
 &\multicolumn{4}{c||}{$\phantom{\Bigl[}B_{V,\,C_A  }^{{\rm DUR}    }(y_{cut})$}\\ \hline
0.01&
 $443 \pm 1.6$ & $329.9 \pm 0.5$ & $181.6 \pm 0.2$ & $71.09 \pm 0.07$ \\                                                                                                   
0.05&
 $116 \pm 0.3$ & $97.26 \pm 0.22$ & $52.4 \pm 0.12$ & $12.98 \pm 0.05$ \\                                                                                                  
0.10&
 $45.25 \pm 0.2$ & $39.53 \pm 0.2$ & $22.5 \pm 0.09$ & $3.62 \pm 0.03$ \\                                                                                                  
0.15&
 $20.4 \pm 0.18$ & $17.82 \pm 0.13$ & $10.37 \pm 0.08$ & $1.16 \pm 0.026$ \\                                                                                               
0.20&
 $8.7 \pm 0.12$ & $7.61 \pm 0.1$ & $4.63 \pm 0.05$ & $0.41 \pm 0.02$ \\                                                                                                    
 \hline
 &\multicolumn{4}{c||}{$\phantom{\Bigl[}B_{V,\,C_F  }^{{\rm DUR}    }(y_{cut})$}\\
\hline
 \hline
 \end{tabular}
 \caption{The DURHAM clustering algorithm.                  }
 \label{DU}
 \end{center}
 \end{table}
\clearpage

\end{document}